\documentclass[onecolumn,journal,draftcls]{IEEEtran}
\usepackage[latin9]{inputenc}
\usepackage{amsmath}
\usepackage{stmaryrd}
\usepackage{graphicx}
\usepackage{epstopdf}
\usepackage[unicode=true,
 bookmarks=true,bookmarksnumbered=true,bookmarksopen=true,bookmarksopenlevel=1,
 breaklinks=false,pdfborder={0 0 0},pdfborderstyle={},backref=false,colorlinks=false]
 {hyperref}
\hypersetup{pdftitle={Your Title},
 pdfauthor={Your Name},
 pdfpagelayout=OneColumn, pdfnewwindow=true, pdfstartview=XYZ, plainpages=false}

\makeatletter

\providecommand{\tabularnewline}{\\}

\let\oldforeign@language\foreign@language
\DeclareRobustCommand{\foreign@language}[1]{%
  \lowercase{\oldforeign@language{#1}}}

\usepackage[caption=false,font=footnotesize]{subfig}

\makeatother

\begin{document}
\title{Low-complexity graph-based traveling wave models
for HVDC grids with hybrid transmission lines: 
Application to fault identification}
\author{Paul~Verrax,~Nathan~Alglave,~ Alberto~Bertinato,\\
Michel~Kieffer,~\IEEEmembership{Senior~Member,~IEEE,}~and~Bertrand~Raison,~\IEEEmembership{Senior~Member,~IEEE}
\thanks{This work was carried out at the SuperGrid Institute, an institute
for the energetic transition (ITE). It is supported by the French
government under the frame of \textquotedblleft Investissements d\textquoteright avenir\textquotedblright{}
program with grant reference number ANE-ITE-002-01.}\thanks{P~Verrax, N~Alglave, and A Bertinato are with the Supergrid Institute,
Villeurbanne, France, e-mail: \protect\href{http://paul.verrax@supergrid-institute.com}{paul.verrax@supergrid-institute.com}.}\thanks{M~Kieffer is with the L2S, Univ Paris-Sud, CNRS, CentraleSupelec,
Univ Paris-Saclay, Gif-sur-Yvette, France, e-mail: \protect\href{http://michel.kieffer@l2s.centralesupelec.fr}{michel.kieffer@l2s.centralesupelec.fr}.}\thanks{B~Raison is with the Univ. Grenoble Alpes, CNRS, Grenoble INP{*},
G2Elab, 38000 Grenoble, France ({*} Institute of Engineering Univ.
Grenoble Alpes), e-mail: \protect\href{http://Bertrand.Raison@univ-grenoble-alpes.fr}{Bertrand.Raison@univ-grenoble-alpes.fr}.}}
\markboth{IEEE Transactions on Smart Grid}{VERRAX \MakeLowercase{\emph{et al.}}: Fast fault identification
in multi-conductor HVDC grids: a fault parameter estimation approach}
\maketitle
\begin{abstract}
The fast protection of meshed HVDC grids requires the modeling of
the transient phenomena affecting the grid after a fault. In the case
of hybrid lines comprising both overhead and underground parts, the
numerous generated traveling waves may be difficult to describe and
evaluate. This paper proposes a representation of the grid as a graph,
allowing to take into account any waves traveling through the grid.
A relatively compact description of the waves is then derived, based
on a combined physical and behavioral modeling approach. The obtained
model depends explicitly on the characteristics of the grid as well
as on the fault parameters. An application of the model to the identification
of the faulty portion of an hybrid line is proposed. The knowledge
of the faulty portion is profitable as faults in overhead lines, generally
temporary, can lead to the reclosing of the line. 
\end{abstract}

\begin{IEEEkeywords}
hybrid lines, fault location, graph theory
\end{IEEEkeywords}

\IEEEpeerreviewmaketitle{}

\section{Introduction }

The integration of renewable energy sources leads to the evolution
of the existing transmission and distribution power grids into a more
interconnected system, known as smart grid \cite{ISSGAN2020}. For
high voltage transmission, direct current (HVDC) technology may outperform
HVAC for long distance interconnections, in particular for underground
or undersea cables \cite{Kalair2016}. In Europe, cables, despite
their more important cost, are more and more preferred to overhead
lines (OHL) due to the difficulty to obtain new right-of-ways for
aerial corridors. It is however possible to upgrade existing HVAC
OHL to HVDC, thus increasing the transmission capacity \cite{Reed2019}.
The recourse to hybrid lines comprising portions of cables and overhead
lines permits a better adaption to different terrains and configurations
(offshore connection, populated areas, existing corridors \textit{etc}.),
see for instance \cite{Vestergaard2019}. Such hybrid lines can be
further integrated into larger Multi-Terminal DC grids (MTDC) to increase
the overall reliability while decreasing the investment costs. 

The protection of MTDC grids against faults remain an open and challenging
topic \cite{Tunnerhoff2019}. The selective clearing of faults by
disconnecting only the affected line is usually the preferred solution
as it allows to operate the healthy parts of the grid continuously.
This requires DC Circuit Breakers (DCCB) at the extremity of each
line. Each DCCB is then controlled by the neighboring relay which
must be able to send the tripping orders as fast as possible (typically
in less than 1 ms). The protection algorithm at the relay must thus
be able to distinguish faults occurring in the protected line (internal
faults) from faults occurring in other parts of the grid (external
faults). Furthermore, in the case of hybrid lines, the identification
of the faulty segment is of interest. While faults affecting cables
are permanent, faults affecting OHL are often temporary and a re-closing
of the line may be attempted. The identification of the faulty segment
within an hybrid line is a difficult task. Many existing approaches
involve distributed sensors at the junction between each portion and
synchronized communication between distant sub-stations, as in \cite{Tzelepis2018}.
On the other hand, single-ended algorithms only require sensors at
the extremity of each line and are thus less sensitive to communication
issues.

For HVAC lines, \cite{Leite2019} proposed a double-ended method using
the arrival times at the two extremities of the hybrid line to estimate
the fault location. The method shows good localization performance
and uncertainty in the line parameters is taken into account to assess
the precision of the estimated fault distance. 

In \cite{Livani2014b}, a Support Vector Machine (SVM) algorithm is
trained to classify the faults of a two segments hybrid line using
single ended data. Voltage and current wavelet energies are used as
inputs for the SVM. Once the faulty section is identified, a wavelet-based
localization technique is applied. As it uses a binary classifier,
this approach is limited to hybrid lines with only two segments. The
SVM has also to be trained with sufficient fault scenarios. For a
Point-to-Point (P2P) hybrid line, \cite{Lewis2016a} showed that the
presence of oscillations in the current evolution after the operation
of the ACCBs is characteristic of a fault in the overhead part of
the line. This kind of approach is not suited for meshed grid where
ACCBs do not operate as primary protection. 

The presence of sensors at the junction is considered in \cite{Tunnerhoff2017}.
A differential protection criterion is then applied to identify the
faulty section, assuming the primary protection is ensured through
the control of full-bridge Modular Multilevel Converter (MMC). Distributed
sensors are also assumed in \cite{Tzelepis2018} in the more general
case of an hybrid link embedded in a MTDC grid. The primary protection
is ensured by a differential current criterion formulated at the junction
points. Localization is then performed using the arrival time difference
at the different sensors, measured through a wavelet transform of
the current.

Model-based approaches representing the transient behavior of the
grid are beneficial as they allow one to exploit the information contained
in the traveling waves appearing after a fault occurrence, see for
instance \cite{Verrax_EPSR_20}. In the case of hybrid lines, however,
the presence of interconnections along the lines render the evaluation
of a large number of traveling waves an arduous task.

This paper proposes a systematic description of the traveling waves
appearing after a fault using a a graph description of the grid. For
each wave, the proposed model combines a physical and a behavioral
part to represent the propagation delays as well as the distortion
due to the ground effects, as detailed in Section~\ref{sec:Systematic-fault-modeling}.
An application example of the model to identify the faulty segment
within a hybrid line is presented in Section~\ref{sec:Faulty-segment-identification}.
Simulation results using a test grid implemented in EMTP-RV \cite{Mahseredjiana1993}
software are presented in Section~\ref{sec:Simulation-results}.

In what follows, Laplace domain or frequency domain variables are
in capital letters, whereas continuous and discrete time domain variables
are in small letters. The convolution is represented by $\otimes$.
$\mathcal{F}$ and $\mathcal{F}^{-1}$ stand for the direct and inverse
Fourier transform. 

\section{Systematic fault modeling \label{sec:Systematic-fault-modeling}}

The section details the modeling of faults affecting hybrid transmission
lines embedded in an MTDC grid with monopolar configuration. The main
notations are introduced in Section~\ref{subsec:Overview-and-notations}.
The TW theory is briefly recalled in Section~\ref{subsec:DC-fault-physical}
and the proposed approach for the systematic modeling of TW is presented.
The behavioral part of the proposed model to account for the soil
resistivity is described in Section~\ref{subsec:Behavioral-modeling-of}. 

\subsection{Overview and notations\label{subsec:Overview-and-notations}}

The considered network is described as an undirected graph $\mathcal{G}=(\mathcal{Q},\mathcal{E})$.
Each vertex $q\in\mathcal{Q}$ represents an interconnection between
two or more line segments. Nodes may correspond to bus-bars or junctions
between overhead line and underground cable segments. Each segment
is represented by an edge $e\in\mathcal{E}$ of the graph. The edge
between the nodes $q_{i}$ and $q_{j}$ is denoted $e_{q_{i}q_{j}}$,
or $e_{i,j}$ to lighten the notations. Since the graph is undirected,
$e_{i,j}=e_{j,i}$. The length of the segment represented by the edge
$e_{i,j}$ is $d_{i,j}$.

We assume that at $t=t_{\text{f}}$ a fault occurs in edge $e_{\text{f}}=e_{i,j}\in\mathcal{E}$
and is modeled as a switch closing in series with the fault resistance
$R_{\text{f}}$ and a constant voltage source. The voltage source
accounts for the collapse of voltage and is set to the opposite of
the pre-fault voltage $V_{\text{bf}}$ at the fault location.

The fault leads to a modification of the graph $\mathcal{G}$. A node
$q_{\text{f}}$ is added to $\mathcal{Q}$ and the faulty edge $e_{\text{f}}=e_{i,j}\in\mathcal{E}$
is replaced by the edges $e_{i,\text{f}}$ and $e_{\text{f},j}$ of
lengths $d_{\text{f},i}$ and $d_{\text{f,}j}$. Formally, the graph
$\mathcal{G}_{\text{f}}=\big(\mathcal{Q}_{\text{f}},\mathcal{E}_{\text{f}}\big)$,
once the fault has occurred, is such that $\mathcal{Q}_{\text{f}}=\mathcal{Q}\cup\left\{ q_{\text{f}}\right\} $
and $\mathcal{E}_{\text{f}}=\mathcal{E}\textbackslash\{e_{\text{f}}\}\cup\big\{ e_{i,\text{f}},e_{j,\text{f}}\big\}$.
The fault can thus be characterized by the vector of fault parameters
$\mathbf{p}=\left(t_{\text{f}},e_{\text{f}},d_{\text{f},i},d_{\text{f},j},R_{\text{f}}\right)$,
where $R_{\text{f}}$ is the fault resistance between the transmission
line and the ground. The two fault distances $d_{\text{f},i}$ and
$d_{\text{f},j}$ are linked through the total length of the line
$d_{\text{f},i}+d_{\text{f},j}=d_{i,j}$, which is known. Thus only
one unknown fault distance is kept, for a fault located on the edge
$e_{i,j}$, the fault distance is arbitrarily defined as
\begin{equation}
d_{\text{f}}=\left\{ \begin{array}{cc}
d_{\text{f},i} & \text{ if }i<j\\
d_{\text{f},j} & \text{ if }j<i
\end{array}\right..\label{eq:df_convention}
\end{equation}

\subsection{DC fault physical modeling\label{subsec:DC-fault-physical}}

This section presents a model describing the transient behavior after
the occurrence of a fault. Preliminaries on traveling waves are first
recalled in Section~\ref{subsec:Traveling-waves}. The proposed systematic
modeling approach to describe any wave traveling from the fault through
the grid is presented in Section~\ref{subsec:Systematic-modeling-of}.

\subsubsection{Traveling waves \label{subsec:Traveling-waves}}

Consider an edge $e_{i,j}$ belonging to a meshed grid such as that
presented in Figure~\ref{fig:test_grid_graph}. The evolution with
time of the current and voltage at a given point of $e_{i,j}$ can
be described using traveling waves, as shown in \cite{Wang2017}.
Along the line, current and voltage satisfy the telegraph equations,
expressed in the Laplace domain as
\begin{align}
\frac{\partial^{2}V}{\partial x^{2}} & =Z(s)Y(s)V(x,s),\\
\frac{\partial^{2}I}{\partial x^{2}} & =Z(s)Y(s)I(x,s),
\end{align}
where $Z(s)=R+sL$ and $Y(s)=G+sC$ are the transfer functions of
the distributed series impedance and shunt admittance, respectively.
In what follows, the distributed parameters $R,L,C$, and $G$ are
considered at a fixed given frequency.

Consider a fault occurring on edge $e_{i,j}$. Two voltage and current
waves $V_{i,1}$ and $V_{j,1}$ travel from the fault location along
the line towards node $q_{i}$ and $q_{j}$, respectively. They undergo
attenuation and distortion described by some propagation function
$H$
\begin{align}
V_{i,1}\left(s,d_{\text{f},i}\right) & =H\left(s,d_{\text{f},i}\right)V_{\text{init}}(s)\label{eq:propagation_i}\\
V_{j,1}\left(s,d_{\text{f},j}\right) & =H\left(s,d_{\text{f},j}\right)V_{\text{init}}(s),\label{eq:propagation_j}
\end{align}
where $V_{\text{init}}$ is the initial surge at fault location. $H$
can be expressed, for a traveled distance $d$ along the line, as
\begin{equation}
H\left(s,d\right)=\exp\left(-\sqrt{Y(s)Z(s)}d\right).\label{eq:PropFunc-1}
\end{equation}
Any voltage traveling wave $V$ has an associated current wave defined
as
\begin{equation}
I(s,d)=Z_{\text{s}}^{-1}(s)V(s,d),\label{eq:Current_relation}
\end{equation}
where
\begin{equation}
Z_{\text{s}}\left(s\right)=\sqrt{Z(s)/Y(s)}\label{eq:CaracImp}
\end{equation}
is the surge (or characteristic) impedance. Similar computations can
be performed for the current.

Getting explicit expressions of the surge impedance and propagation
function requires several approximations. Considering a low-loss approximation,
one has
\[
Z_{\text{s}}\left(s\right)=\sqrt{\frac{R+sL}{G+sC}}\simeq\sqrt{\frac{L}{C}}\left(1+\frac{1}{2}\left(\frac{R}{sL}-\frac{G}{sC}\right)\right).
\]
Further neglecting the shunt admittance $G$, one gets 
\begin{equation}
Z_{\text{s}}\left(s\right)\simeq\sqrt{\frac{L}{C}}\left(1+\frac{1}{2}\frac{R}{sL}\right).\label{eq:Zs_lowloss}
\end{equation}
For the propagation constant $\gamma\left(s\right)=\sqrt{Y(s)Z(s)}$,
the low-loss approximation leads to 
\[
\gamma\left(s\right)\simeq s\sqrt{LC}\left[1+\frac{1}{2}\left(\frac{R}{sL}+\frac{G}{sC}\right)\right]
\]
and neglecting again the shunt admittance $G$, 
\begin{equation}
\gamma\left(s\right)\simeq s\sqrt{LC}\left[1+\frac{1}{2}\frac{R}{sL}\right].\label{eq:gamma_lowloss}
\end{equation}

Considering the lossless approximation, one gets 
\begin{align}
Z_{\text{s}} & \simeq\sqrt{\frac{L}{C}}\label{eq:Zs_lossless}\\
\gamma\left(s\right) & \simeq s\sqrt{LC}\label{eq:gamma_lossless}
\end{align}
In that case, the surge impedance is real and the propagation function
$H\left(s,d\right)=\exp\left(-\gamma\left(s\right)d\right)$ is a
pure delay.

In practice, the lossless approximations appears to be sufficient
in the considered fault localization context. Nevertheless, the characteristic
impedance of underground cables still require a low-loss approximation
to provide results of sufficient accuracy.

When a change of propagation medium occurs (typically at the junction
between a line and a station), the incident wave $V_{\text{f}}$ induces
a transmitted wave $V_{\text{t}}$ and a reflected wave $V_{\text{r}}$.
The associated voltage $V_{\text{tot}}$ at the junction is 
\begin{align}
V_{\text{tot}} & =V_{\text{t}}=V_{\text{f}}+V_{\text{r}}\nonumber \\
 & =\left(1+K\right)V_{\text{f}}\nonumber \\
 & =TV_{\text{f}}.\label{eq:MediaChange}
\end{align}
The transmission and reflection coefficients $T$ and $K$ depend
on the characteristic admittance $Y_{\text{s}}=Z_{\text{s}}^{-1}$
of the media. Consider a node $q$ connected to $n+1$ edges $e_{0},e_{1},\dots e_{n}$,
the reflection coefficient for a wave traveling from edge $e_{0}$,
reflected at node $q$, an traveling backwards $e_{0}$ is
\begin{align}
K_{e_{0}\hookleftarrow q} & =\frac{Y_{e_{0}}-\sum_{\ell=1}^{n}Y_{s,e_{\ell}}}{\sum_{\ell=0}^{n}Y_{s,e_{\ell}}}.\label{eq:refl_coef}
\end{align}
The transmission coefficient from edge $e_{0}$ through node $q$,
$i=1,\dots,n$ is
\begin{equation}
T_{e_{0}\shortrightarrow q}=1+K_{e_{0}\hookleftarrow q}=\frac{2Y_{s,e_{0}}}{\sum_{\ell=0}^{n}Y_{s,e_{\ell}}}.\label{eq:trans_coef}
\end{equation}
Since the fault is modeled as a switch closing at $t=t_{\text{f}}$
in series with the fault resistance $R_{\text{f}}$ and a constant
voltage source of amplitude $V_{\text{bf}}$, the initial surge at
the fault location $V_{\text{init}}$ is thus modeled as
\begin{align}
V_{\text{init}} & =\underbrace{\frac{-1/R_{\text{f}}}{2/Z_{\text{s}}+1/R_{\text{f}}}}_{K_{e_{i,\text{f}}\hookleftarrow e_{\text{f}}}}V_{\text{bf}}\exp\left(-st_{\text{f}}\right)\label{eq:V_init_expr}
\end{align}
which can also be expressed using the reflection coefficient from
the line to the fault $K_{e_{i,\text{f}}\hookleftarrow e_{\text{f}}}.$
The voltage at the fault location just before the fault occurrence
$V_{\text{bf}}$ can be approximated by the measured pre-fault voltage
at the relay $q$.

For reflection and transmission at sub-stations comprising MMCs, we
adopt an RLC equivalent model \cite{Leterme2014}, which is valid
before the blocking of the station
\begin{equation}
Z_{\text{mmc}}\left(s\right)=R_{\text{mmc}}+sL_{\text{mmc}}+\frac{1}{sC_{\text{mmc}}}.\label{eq:MMCmodel}
\end{equation}
The propagation equations \eqref{eq:propagation_i} and \eqref{eq:propagation_j}
combined with the reflection and transmission equations \eqref{eq:MediaChange},
\eqref{eq:refl_coef}, and \eqref{eq:trans_coef} allow one to model
any particular wave traveling from the fault to the grid. Nevertheless,
a faulty grid comprising hybrid lines will host many reflected waves
due to the multiple junctions. A systematic approach describing these
traveling waves is thus required and detailed in Section~\ref{subsec:Systematic-modeling-of}.

\subsubsection{Systematic model of traveling waves within a grid \label{subsec:Systematic-modeling-of}}

Consider a node $q_{\text{s}}\in\mathcal{Q}_{\text{f}}$ at which
voltage and current are observed. This node may, for instance, connect
multiple transmission lines to a converter station. The aim in what
follows is to propose a \emph{physical} model of the TWs caused by
the fault and reaching $q_{\text{s}}$. A TW is entirely determined
by its \textit{path}, \textit{i.e.}, the sequence of nodes it has
traversed. Formally, all possible paths from $q_{\text{f}}$ to $q_{\text{s}}$
can be defined as
\begin{multline}
\mathcal{P}_{q_{\text{f}}\shortrightarrow q_{\text{s}}}=\left\{ \left(q_{n_{1}},..,q_{n_{m}}\right)|\right.\\
\left.q_{n_{1}}=q_{\text{f}},q_{n_{m}}=q_{\text{s}},\left(q_{n_{i}},q_{n_{i+1}}\right)\in\mathcal{E}_{\text{f}},m>1\right\} \label{eq:all_paths}
\end{multline}
A path $\pi\in\mathcal{P}_{q_{\text{f}}\shortrightarrow q_{\text{s}}}$
may comprise the same node several times, including the faulty node
$q_{\text{f}}$ and the observation node $q_{\text{s}}$. Due to the
reflections occurring at the junctions, a TW is indeed likely to pass
several times via the same nodes. Considering the lossless approximation
and constant distributed parameters, as in Section~\ref{subsec:Traveling-waves}
when a wave travels on an edge, only the propagation delay has to
be taken into account. Consequently, when modeling traveling waves,
one has to account for
\begin{itemize}
\item the different delays due to the propagation along the edges,
\item the effect of junctions on the incident wave.
\end{itemize}
Consider a path $\pi=\left(q_{n_{1}},..,q_{n_{m}}\right)\in\mathcal{P}_{q_{\text{f}}\shortrightarrow q_{\text{s}}}$
traveled by a given wave. The total propagation delay along $\pi$
is
\begin{equation}
\tau_{\pi}(d_{\text{f}})=\sum_{i=1}^{m}\Delta t_{n_{i},n_{i+1}}=\sum_{i=1}^{m-1}\frac{d_{n_{i},n_{i+1}}}{c_{n_{i},n_{i+1}}}\label{eq:prop_delay}
\end{equation}
where $c_{n_{i},n_{i+1}}$ is the wave propagation speed along the
edge $\left(q_{n_{i}},q_{n_{i+1}}\right)$, depending on the propagation
medium. The total delay thus depends on the fault distances $d_{\text{f},i}$
or $d_{\text{f,}j}$ as at least the first traversed edge is necessarily
connected to the fault.

At each junction along $\pi$, the voltage wave is subject to a transmission
and a reflection. The resulting coefficient depends on the propagation
direction before and after the junction. Consequently, the impact
of reflections and transmissions at junctions is described by
\begin{equation}
V_{\pi,\text{j}}(s,t_{\text{f}},R_{\text{f}})=\prod_{i=1}^{m}J_{e_{n_{i-1},n_{i}}\shortrightarrow q_{n_{i}}}(s,R_{\text{f}})\frac{\exp(-t_{\text{f}}s)}{s}V_{\text{bf}}\label{eq:junction_interact}
\end{equation}
where $J$ is either a reflection \eqref{eq:refl_coef} or transmission
\eqref{eq:trans_coef} coefficient 
\[
J_{e_{n_{i-1},n_{i}}\shortrightarrow q_{n_{i}}}=\left\{ \begin{array}{cc}
T_{e_{n_{i-1},n_{i}}\shortrightarrow q_{n_{i}}} & \text{ if }n_{i-1}\neq n_{i+1}\\
K_{e_{n_{i-1},n_{i}}\hookleftarrow q_{n_{i}}} & \text{ if }n_{i-1}=n_{i+1}
\end{array}\right.
\]
for $i=2,\dots,m-1$. The first term in the product \eqref{eq:junction_interact}
accounts for the initial surge at the fault location $J_{e_{n_{0},n_{i}}\shortrightarrow q_{n_{1}}}=K_{q_{_{n_{1}},},q_{_{n_{2}},}\hookleftarrow q_{\text{f}}}\left(R_{\text{f}}\right)$
and depends thus on the fault resistance, see \eqref{eq:V_init_expr}.
The voltage at node $q_{\text{s}}$ due to the arrival of an incident
wave corresponds to the transmitted wave to the node $q_{\text{s}}$
\eqref{eq:MediaChange}. This transmission coefficient is thus included
as the last term in the product \eqref{eq:junction_interact}, hence
$J_{e_{n_{m-1},n_{m}}\shortrightarrow q_{m}}=T_{e_{n_{m-1},n_{m}}\shortrightarrow q_{\text{s}}}$.
Consequently, for a given path $\pi$, considering the propagation
delay \eqref{eq:prop_delay} and the transmissions and reflections
occurring along $\pi$ via \eqref{eq:junction_interact}, one gets
the following \textit{physical model}
\begin{equation}
V_{\pi}^{0}\left(s,\mathbf{p}\right)=\exp\left(-\tau_{\pi}(d_{\text{f}})s\right)V_{\pi,\text{j}}(s,t_{\text{f}}R_{\text{f}})\label{eq:physical_model}
\end{equation}
of the voltage caused by a TW along $\pi$.

\begin{figure}[tbh]
\begin{centering}
\includegraphics[width=0.5\textwidth]{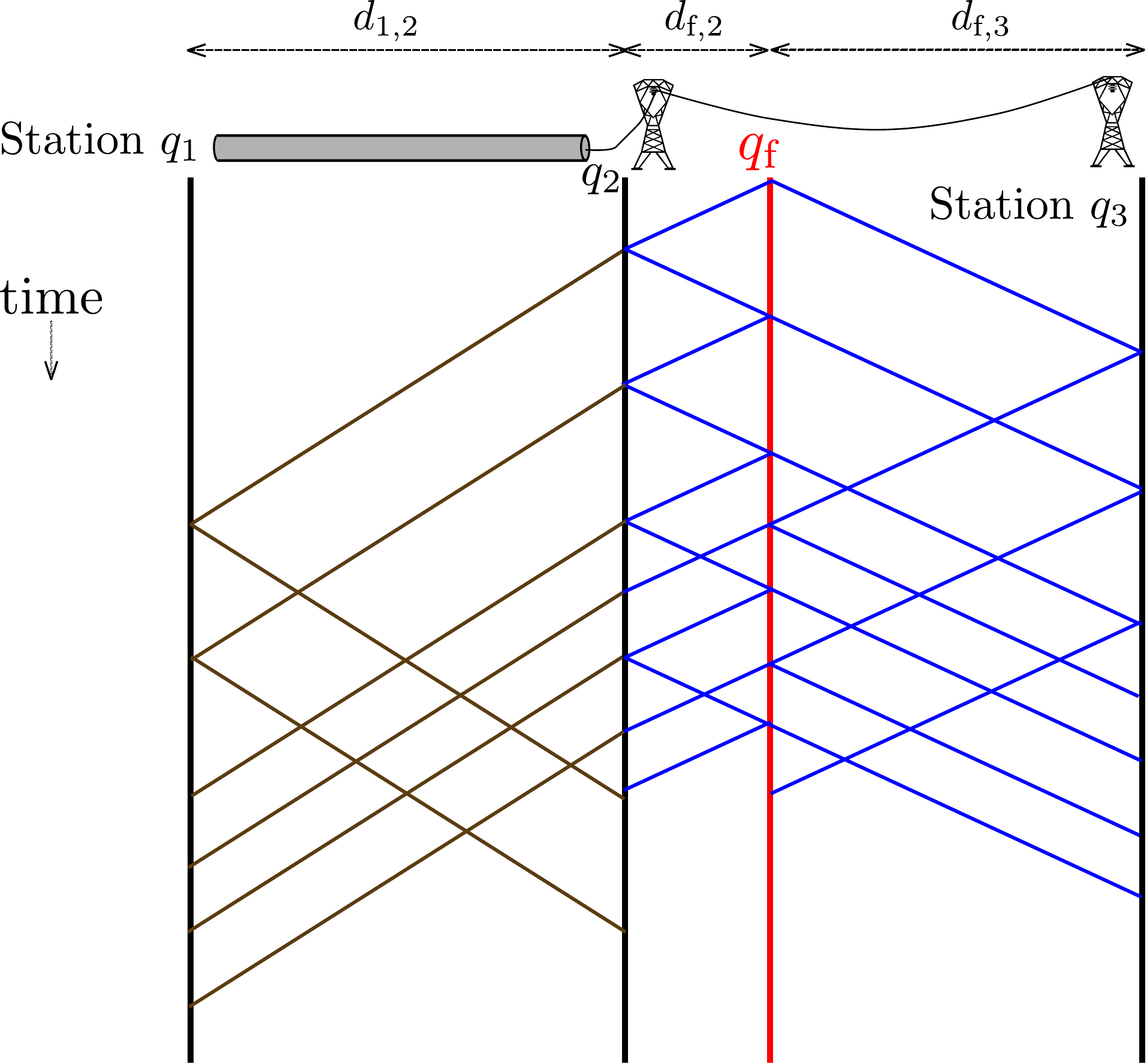}
\par\end{centering}
\caption{Example of Bewley lattice diagram for a hybrid point-to-point link
when a fault occurs as $q_{\text{f}}$ located in an overhead portion
of the line.\label{fig:Example-of-Bewley's}}
\end{figure}

The different paths taken by the TW can be represented via a Bewley
lattice diagram. Figure~\ref{fig:Example-of-Bewley's} illustrates
such diagram on a point-to-point link consisting of an OHL and a cable
segment. Even in this relatively simple case, the presence of the
OHL-cable junction creates a large number of reflected TWs. The propagation
speed of the TWs in the underground part is slower than in the overhead
line.

The waveform in the time domain is obtained through the inverse Fourier
transform.
\begin{align}
v_{\pi}^{0}(t,\mathbf{p}) & =\mathcal{F}^{-1}\left\{ \left.V_{\pi}^{0}(s,\mathbf{p})\right|_{s=j\omega}\right\} \nonumber \\
 & =\mathcal{F}^{-1}\left\{ V_{\pi,\text{j}}(\omega,R_{\text{f}})\right\} \otimes\delta_{\tau_{\pi}\left(d_{\text{f}}\right)}\left(t\right)\nonumber \\
 & =v_{\pi,\text{j}}(t,R_{\text{f}})\otimes\delta_{\tau_{\pi}}\left(d_{\text{f}},t\right)\nonumber \\
 & =v_{\pi,\text{j}}(t-\tau_{\pi}\left(d_{\text{f}}\right),R_{\text{f}})\label{eq:phy_model_temp}
\end{align}
where $\delta_{\tau_{\pi}\left(d_{\text{f}}\right)}\left(t\right)=\delta_{0}\left(t-\tau_{\pi}\left(d_{\text{f}}\right)\right)$
is the Dirac distribution corresponding to the propagation delay $\tau_{\pi}$
along the path. In practice, $\mathcal{F}^{-1}$ is computed numerically
using the inverse discrete Fourier transform. Considering a sampling
period $T_{\text{s}}$, the obtained discrete-time voltage model at
time $kT_{\text{s}}$ is thus written as $v_{\pi}^{0}\left(k,\mathbf{p}\right)$.

\subsection{Behavioral modeling of the ground effects\label{subsec:Behavioral-modeling-of}}

The physical model developed in Sections~\ref{subsec:Traveling-waves}
and \ref{subsec:Systematic-modeling-of} assumes the distributed line
parameters as independent of the frequency. With this approximation,
distortions of the waves cannot be described. In particular, the soil
resistivity effects for the OHL portions as well as the screen resistance
for the cable portions are not taken into account.

To account for such effects, a behavioral model is proposed in this
section. It extends to hybrid lines a model previously introduced
for overhead lines in \cite{Verrax_EPSR_20}. From the geometry of
the transmission lines and the characteristics of the conductors,
the response for a voltage step propagating along a given edge $e$
can be obtained using EMT simulation software. In particular, the
step response depends on the length of the considered segment $d_{e}$
and on the value of the soil resistivity $\rho_{e}$. The latter is
considered as a known constant characteristic of the considered line.
The propagation delay is removed from the step responses as it is
already accounted for by the propagation constant \eqref{eq:gamma_lossless}.
\begin{figure}[tbh]
\begin{centering}
\includegraphics[width=0.5\columnwidth]{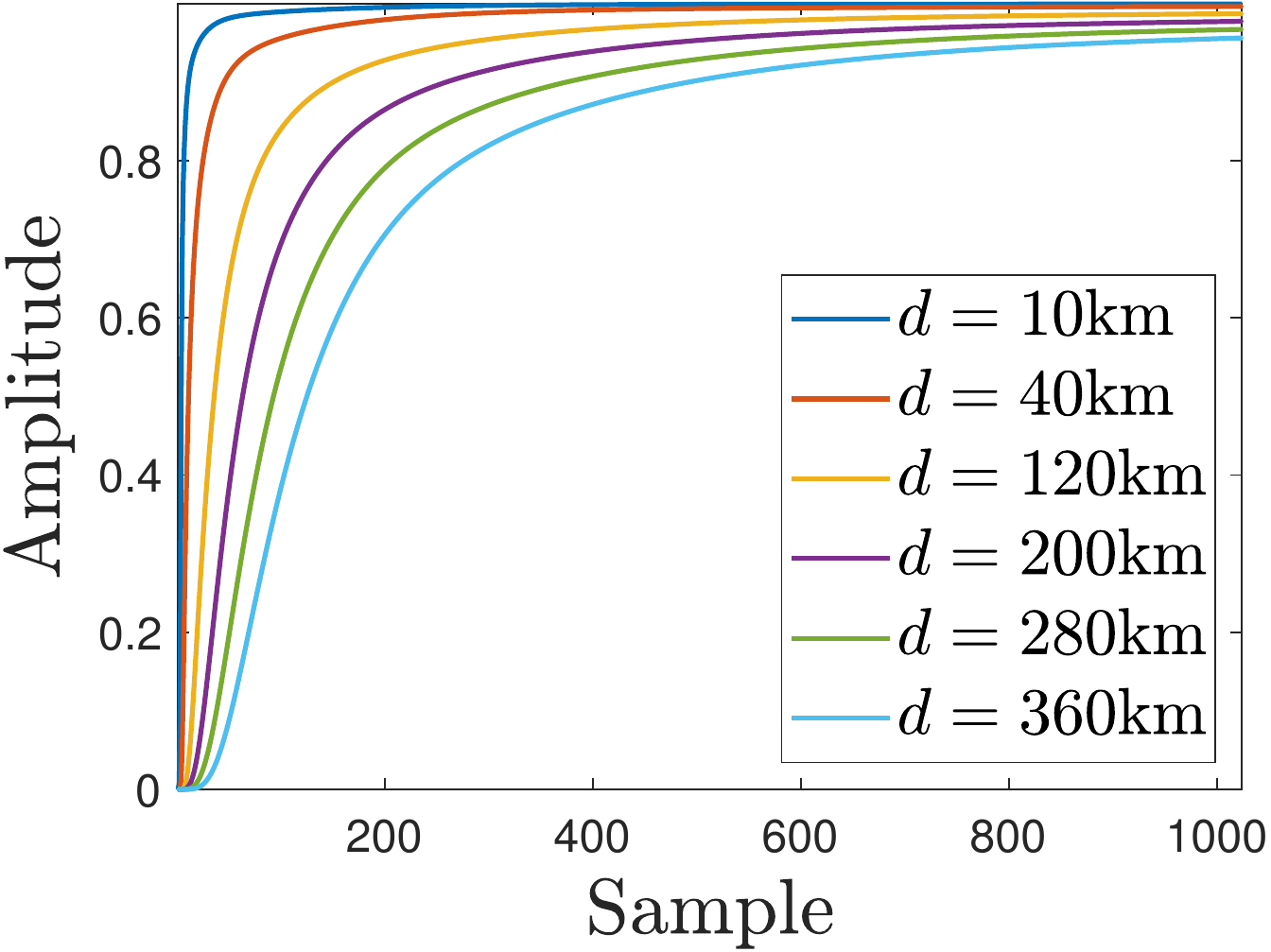}\includegraphics[width=0.5\columnwidth]{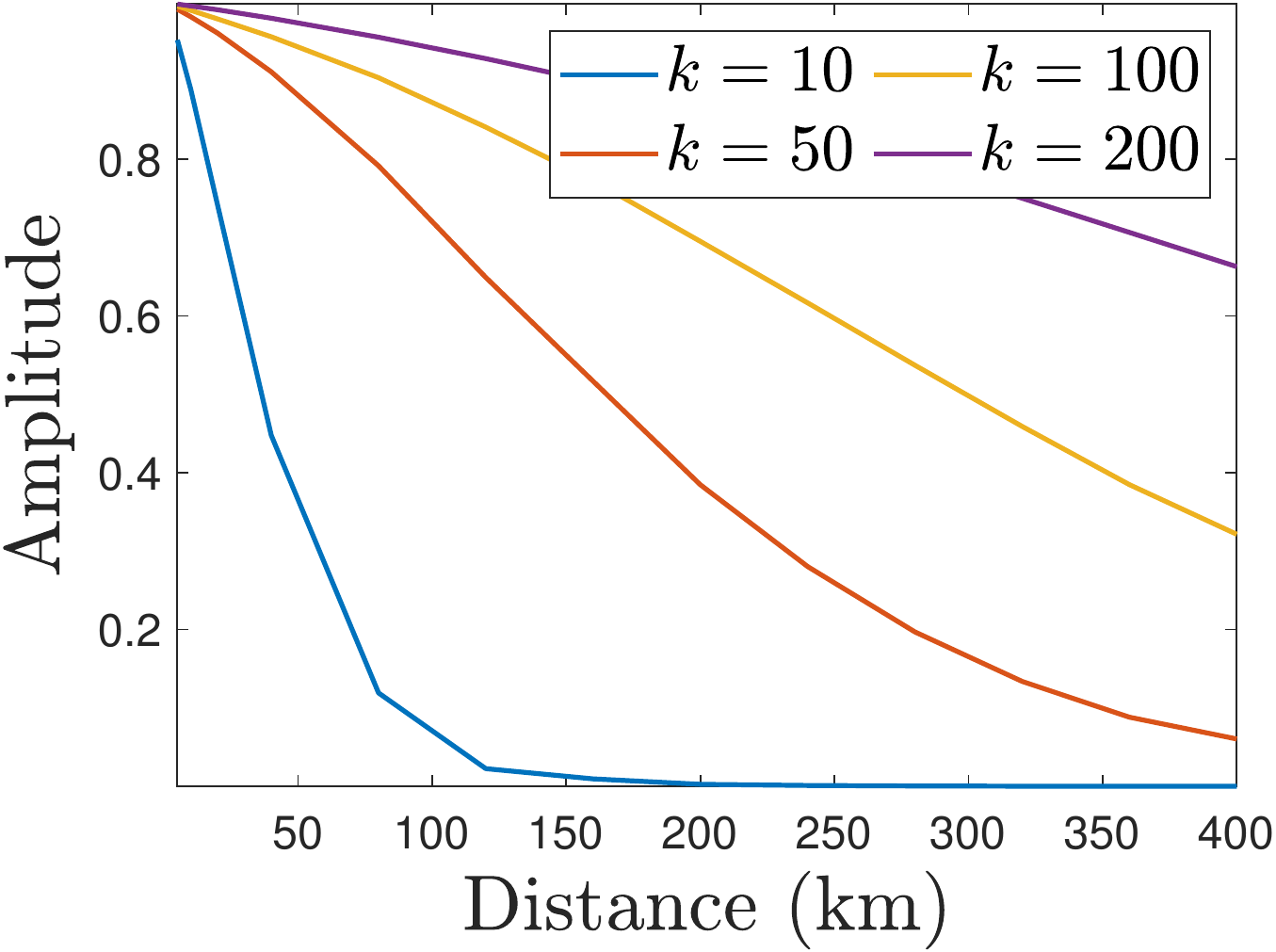}
\par\end{centering}
\begin{centering}
\includegraphics[width=0.5\columnwidth]{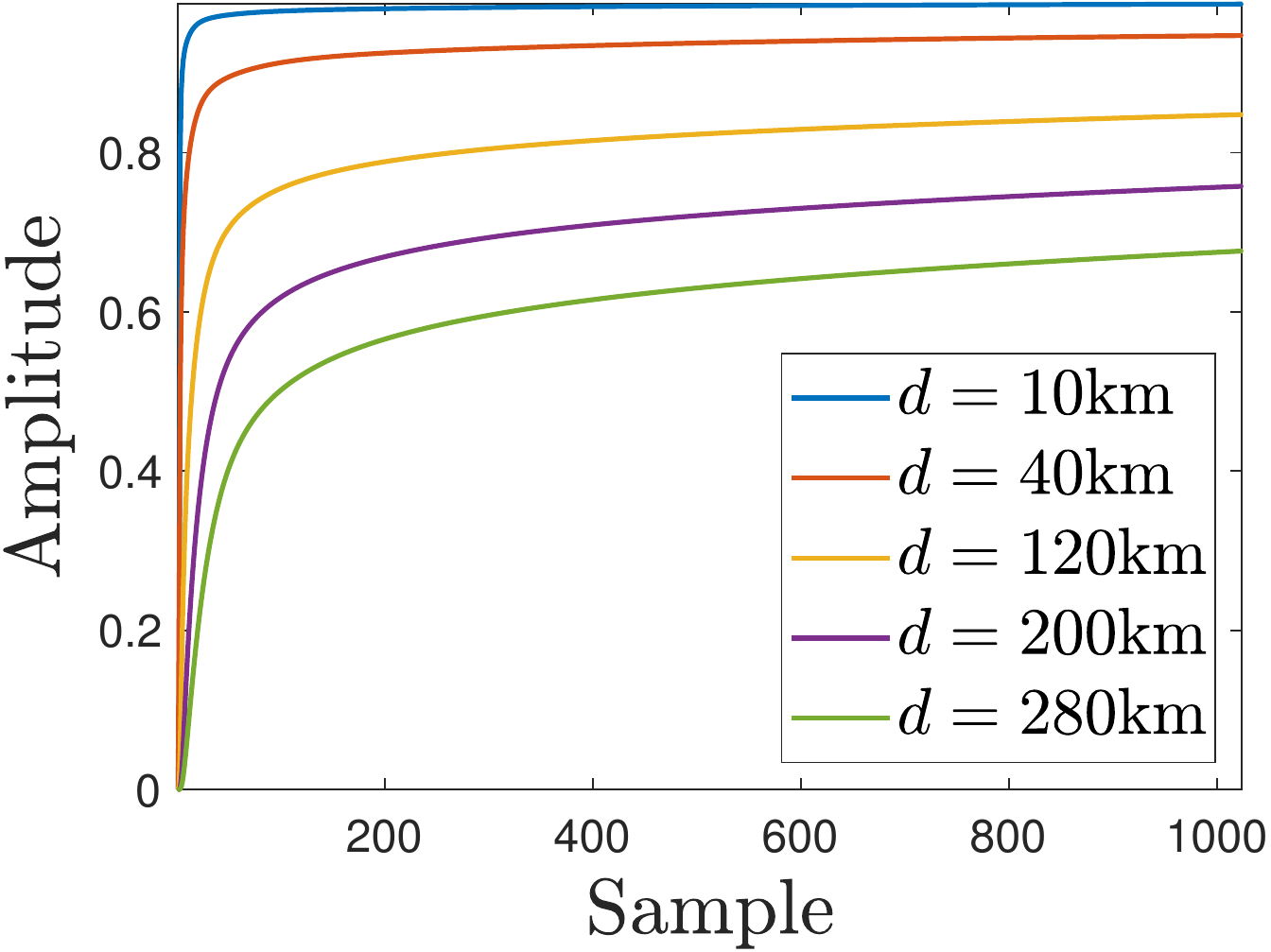}\includegraphics[width=0.5\columnwidth]{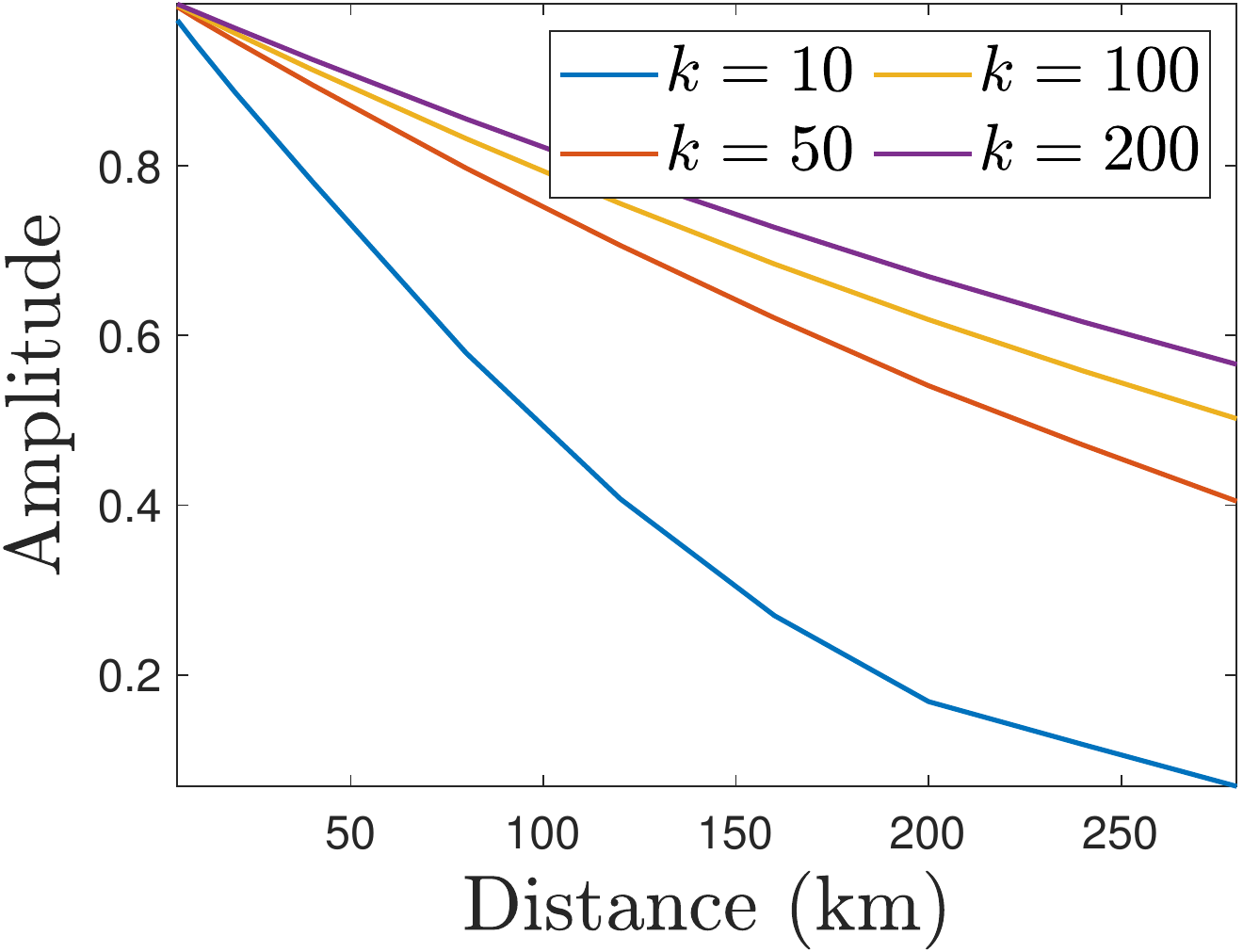}
\par\end{centering}
\caption{Unit step response for overhead lines (top-left) and cables (bottom-left)
of different lengths; The variation of amplitude with the distance
for specific sample points is detailed for OHL (top-right) and cables
(bottom-right); The sampling frequency is $f_{\text{s}}=1\,\text{MHz}$.
\label{fig:Unit-step-response}}
\end{figure}

Assume that a set of known step responses $u_{d,\rho}(k)$ for various
edge lengths $\left\{ d_{1},d_{2},\dots d_{n}\right\} $ is available
for a given conductor and line geometry, as presented in Figure~\ref{fig:Unit-step-response}.
The different step responses have smooth variations with respect to
the line length. Thus, to obtain a step response $u_{d}$ for any
length $d$ such that $d_{i}<d<d_{i+1}$, $i=1,\dots,n-1$ , we propose
an interpolation using the step responses obtained for fault distances
$d_{i}$ and $d_{i+1}$
\begin{equation}
u_{d}(k)=\frac{u_{d_{i+1}}(k)-u_{d_{i}}(k)}{(d_{i+1}-d_{i})}(d-d_{i})+u_{d_{i}}(k).\label{eq:step_d_interp}
\end{equation}
The step response of a given edge $e$ is differentiated to obtain
the impulse response $h_{e}$
\begin{equation}
h_{e}(k)=\frac{u_{d,\rho}(k+1)-u_{d,\rho}\left(k\right)}{T_{\text{s}}}.\label{eq:impulse_he}
\end{equation}

If the step response for the soil resistivity $\rho_{e}$ of the edge
$e$ is unknown, it can be interpolated from the step responses at
known soil resistivities $\rho_{1},\dots,\rho_{m}$ similarly to \eqref{eq:step_d_interp}.

\begin{figure}[tbh]
\centering
\includegraphics[width=0.5\textwidth]{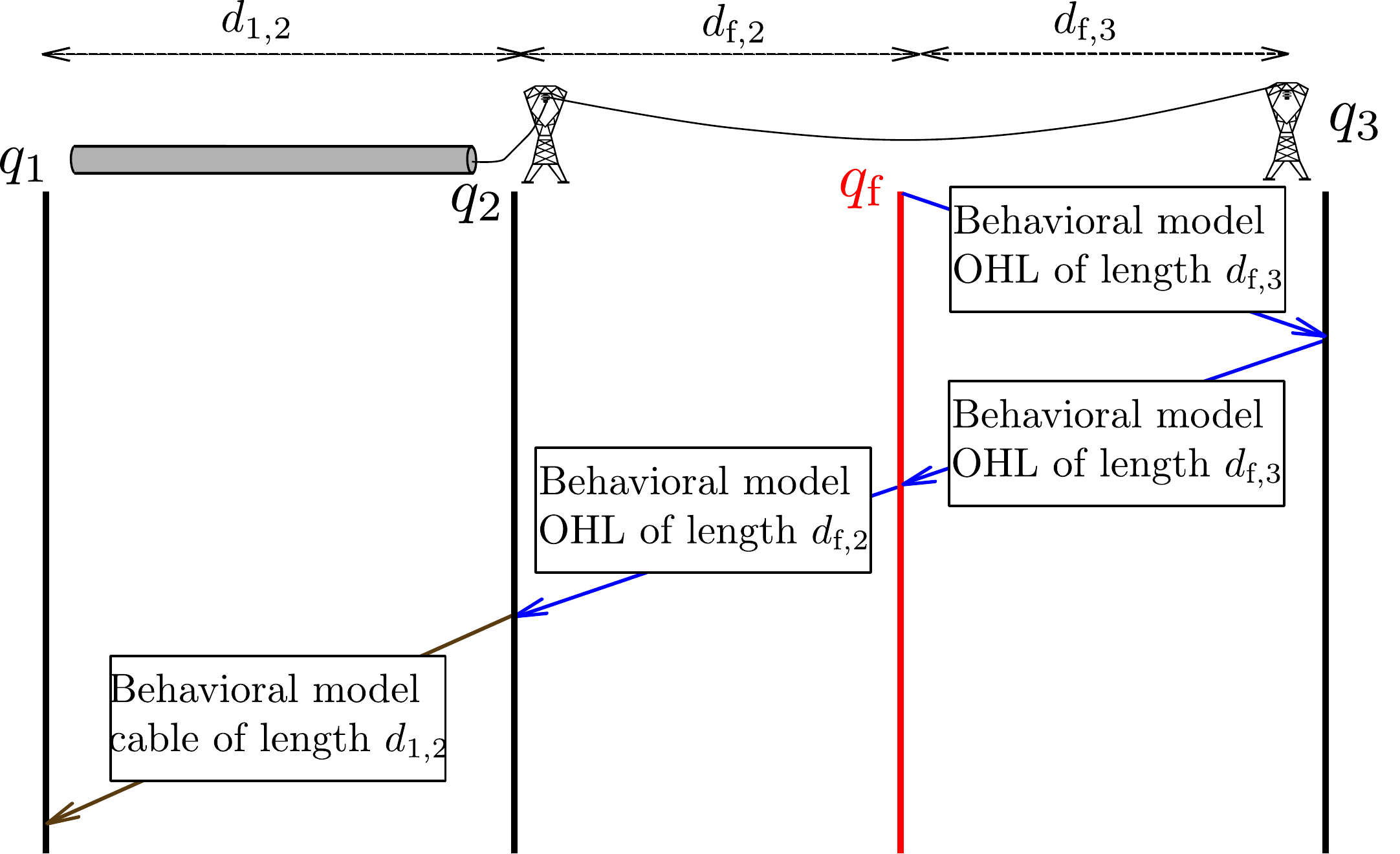}

\caption{Example of cascaded behavioral model to take into account segmented
transmission lines\label{fig:Example-of-cascaded}}
\end{figure}

The evolution of a wave traveling through an edge $e$ of length $d_{e}$
is obtained as the output of the finite impulse response filter excited
by the output of the physical model \eqref{eq:physical_model} of
the edge $e$ 
\begin{equation}
v_{e}^{\text{m}}\text{\ensuremath{\left(k,\mathbf{p}\right)}}=h_{e}\left(k,d_{e}\right)\otimes v_{e}^{0}\left(k,\mathbf{p}\right).\label{eq:model_vme}
\end{equation}
For a wave traveling through a path $\pi\in\mathcal{P}$ comprising
several edges, the total voltage evolution at node $q_{\text{s}}$
is obtained by cascading the impulse responses of the different edges,
see for example Figure~\ref{fig:Example-of-cascaded}
\begin{align}
v_{\pi}^{\text{m}} & \left(k,\mathbf{p}\right)=\underbrace{\bigotimes_{e\in\pi}h_{e}\left(k,d_{e}\right)}_{=h_{\pi}\left(k,d_{\text{f}}\right)}\otimes v_{\pi}^{0}(k,\mathbf{p})\label{eq:model_v_pi_m}
\end{align}

 The global model of the voltage at the node of interest $q_{\text{s}}$
has thus to gather all possible traveling waves between the faulty
node $q_{\text{f}}$ and $q_{\text{s}}$
\begin{equation}
v_{q_{\text{s}}}^{\text{m}}\left(\mathbf{p},k\right)=\sum_{\pi\in\mathcal{P}_{q_{\text{f}}\shortrightarrow q_{\text{s}}}}v_{\pi}^{\text{m}}\left(\mathbf{p},k\right).\label{eq:all_paths_model}
\end{equation}

When considering a finite observation window of duration $\tau_{\text{max}}$
after the occurrence of a fault, only a finite number of traveling
waves may reach the node $q_{\text{s}}$ within this time observation
window. This reduces the set of paths to consider for simulation
\begin{equation}
\mathcal{P}_{q_{\text{f}}\shortrightarrow q_{\text{s}},\tau_{\text{max}}}=\left\{ \pi\in\mathcal{P}_{q_{\text{f}}\shortrightarrow q_{\text{s}}}|\tau_{\pi}<\tau_{\text{max}}\right\} .\label{eq:all_paths_tmax}
\end{equation}
An alternative approach to limit the computational complexity is to
simulate a maximum number $n_{\text{max}}$ of TWs and to consider
as many paths.

\section{Faulty segment identification \label{sec:Faulty-segment-identification}}

This section describes an extension of the fault identification algorithm
presented in \cite{Verrax_EPSR_20} in the case of overhead lines
only. The estimation of the fault parameters is first summarized in
Section~\ref{subsec:Fault-parameter-estimation}. For hybrid lines,
the fault identification algorithm must determine whether the line
under protection is faulty and assert which of the segments is affected
by the fault. Thus leads to a multiple hypothesisis approach, as presented
in Section~\ref{subsec:Fault-identification}.

\subsection{Fault parameter estimation \label{subsec:Fault-parameter-estimation}}

Consider a relay at some node $q$ monitoring a line $L$ described
by $m$ edges $\left(e_{1,2},\dots,e_{m-1,m}\right)$ of lengths $\left(d_{1,2},\dots,d_{i,i+1},\dots,d_{m-1,m}\right)$.
A fault occurs at time $t_{\text{f}}$ in an edge $e_{\text{f}}=e_{i,i+1}$
within line $L$. The vector of the fault parameters to be estimated
is $\mathbf{p}=\left(t_{\text{f}},d_{\text{f}},R_{\text{f}},e_{\text{f}}\right)$.
The time at which the first wave induced by the fault reaches node
$q$ is related to $t_{\text{f}}$ as 
\begin{equation}
t_{\text{d},q}=t_{\text{f}}+\sum_{k=1}^{i-1}\frac{d_{k}}{c_{k}}+\frac{d_{\text{f}}}{c_{i,i+1}}.\label{eq:t_detec}
\end{equation}
We assume that the detection time $t_{\text{d},q}=k{}_{\text{d},q}T_{\text{s}}$
can be accurately measured at the station $q$.

The parametric model developed in Section~\ref{sec:Systematic-fault-modeling}
is employed to estimate the fault parameters. As, this model requires
the faulty edge to be fixed, several hypotheses related to the faulty
edge $e_{\text{\text{f}}}$ have to be considered in parallel to estimate
the fault parameters $\mathbf{p}=\left(d_{\text{f}},R_{\text{f}},e_{\text{f}}\right)$,
where $t_{\text{f}}$ is deduced from $t_{\text{d},q}$ and $d_{\text{f}}$
using \eqref{eq:t_detec}. Under hypothesis $\mathcal{H}_{\ell}$,
the fault is assumed to be located in the edge $e_{\ell,\ell+1}\in\mathcal{E_{\text{f}}}$,
$\ell=1,\dots,m-1$ and the vector of parameters to be estimated boils
down to $\mathbf{p}_{\ell}=\left(d_{\ell,\text{f}},R_{\text{f}}\right)$.

The algorithm evaluates a maximum likelihood estimate $\mathbf{\widehat{p}}_{\ell}$
of the vector of fault parameters $\mathbf{p}_{\ell}$ using the voltage
and current measurements $\left(v_{q}\left(k\right),i_{q}\left(k\right)\right)$
and the model $\left(v_{q,\ell}^{\text{m}}\left(\mathbf{p},k\right),i_{q,\ell}^{\text{m}}\left(\mathbf{p},k\right)\right)$
associated to the hypothesis $\mathcal{H}_{\ell}$. Considering that
the voltage and current measurement noises are realizations of independent
and identically distributed zero-mean Gaussian variables of respective
variances $\sigma_{v}^{2}$ and $\sigma_{i}^{2}$, when $n$ measurements
are available, evaluating $\mathbf{\widehat{p}}_{\ell}$ amounts to
minimizing the following cost function \cite{Walter1997}
\begin{multline}
c_{\ell}^{\left(n\right)}\left(\mathbf{p}\right)=\frac{1}{\sigma_{v}^{2}}\sum_{k=k_{\text{d},q}}^{k_{\text{d},q}+n-1}\left(v_{q,\ell}^{\text{m}}\left(\mathbf{p},k\right)-v_{q}\left(k\right)\right)^{2}+\\
\frac{1}{\sigma_{i}^{2}}\sum_{k=k_{\text{d},q}}^{k_{\text{d},q}+n-1}\left(i_{q,\ell}^{\text{m}}\left(\mathbf{p},k\right)-i_{q}\left(k\right)\right)^{2}.\label{eq:cost_func}
\end{multline}
The algorithm evaluates iteratively an estimate $\mathbf{\widehat{p}}_{\ell}$
of the fault parameters. The estimation algorithm starts when an abnormal
behavior is detected at the relay and the estimate 
\[
\mathbf{\widehat{p}}_{\ell}^{\left(n\right)}=\arg\min_{\mathbf{p}}c_{\ell}^{\left(n\right)}\left(\mathbf{p}\right)
\]
is updated when $\Delta n$ new measurements are available. This minimization
is performed considering Levenberg-Marquadt's algorithm, which requires
an evaluation of the partial derivatives of $c_{\ell}^{\left(n\right)}\left(\mathbf{p}\right)$
with respect to $d_{\text{f}}$ and $R_{\text{f}}$. This may be done
by finite differences, leading to a computational cost which is twice
that of evaluating the cost. Appendices~\ref{subsec:Partial-derivative-dist}
and \ref{subsec:Partial-derivative-resist} detail an explicit evaluation
of these partial derivatives. Several simplifications are possible,
which reduces the complexity of the computations.

\subsection{Fault identification \label{subsec:Fault-identification}}

For a given hypothesis $\mathcal{H}_{\ell}$, the algorithm determines
after each iteration whether $\mathbf{\widehat{p}}_{\ell}^{\left(n\right)}$
is a satisfying estimate of the fault parameters, \emph{i.e.}, if
it is compliant with $\mathcal{H}_{\ell}$ regarding the geometry
of the edge and if the estimate has been obtained with a sufficient
level of confidence. Two tests are employed to confirm or reject the
hypothesis $\mathcal{H}_{\ell}$ that the segment $e_{\ell}$ is faulty.

First, a \emph{validity} test determines whether or not $\mathbf{\widehat{p}}_{\ell}^{\left(n\right)}$
is included in some domain of interest. For instance, the estimated
fault distance should be less than the total length of the assumed
faulty edge $e_{\ell,\ell+1}$. This domain of interest thus depends
on the assumed faulty edge as well as the type of segment, as for
instance fault resistances in cables are much lower than in overhead
lines. 

Second, an \textit{accuracy} test determines whether the area of the
$\alpha$ confidence region of the estimated parameters, $\mathcal{R}^{(\alpha)}(\mathbf{\widehat{p}}_{\ell}^{\left(n\right)})$,
is less than a threshold $t_{\alpha}$. The $\alpha=95\%$ confidence
region is considered. The confidence region is computed based on the
Fisher information matrix, assuming usual statistical properties such
as normal independent distribution of the measurement noises, \cite{Walter1997}.
If both tests are satisfied, the fault is deemed to potentially affect
edge $e_{\ell,\ell+1}$. Otherwise, the algorithm waits until $\Delta n$
additional measurements are available to update $\mathbf{\widehat{p}}_{\ell}^{\left(n\right)}$
and $\mathcal{R}^{(\alpha)}(\mathbf{\widehat{p}}_{\ell}^{\left(n\right)})$.
Once enough measurements have been made available without allowing
to conclude, the hypothesis $\mathcal{H}_{\ell}$ is rejected.

After considering $n$ measurements, several edges $e_{\ell,\ell+1}$
may be deemed to be affected by a fault. Assuming there is a single
fault, the algorithm determines which segment is actually faulty by
considering the hypothesis with the smallest cost \eqref{eq:cost_func}
\[
\widehat{e}_{\text{f}}=\arg\min_{e_{\ell,\ell+1}}\left\{ c^{\left(n\right)}(\mathbf{p},e_{\ell,\ell+1})|\text{fault is identified on }e_{\ell,\ell+1}\right\} .
\]

\section{Simulation results\label{sec:Simulation-results} }

This section presents the results of the fault identification algorithm
implementing the hybrid model considering the EMT software EMTP-RV
\cite{Mahseredjian2007} to simulate the behavior of a grid affected
by a fault. The test grid is described in Section~\ref{subsec:Test-grid}.
The model proposed in Section~\ref{sec:Systematic-fault-modeling}
is implemented in Matlab and compared against EMT simulations in Section~\ref{subsec:Modeling-results}.
An illustrative example of the fault identification approach is detailed
in Section~\ref{subsec:Illustrative-example}.

\subsection{Test grid\label{subsec:Test-grid}}

\begin{figure}[tbh]
\begin{centering}
\includegraphics[width=0.5\textwidth]{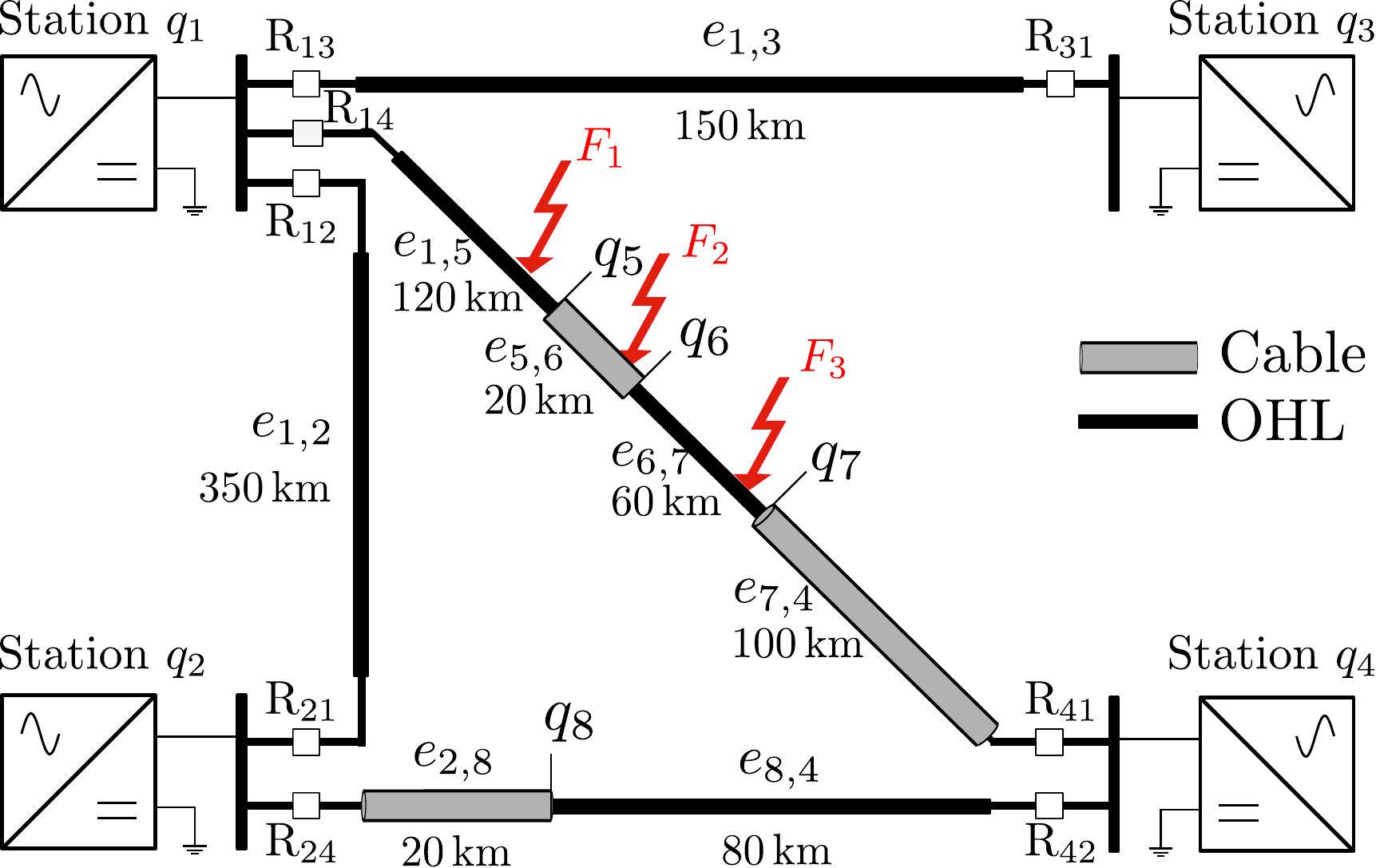}
\par\end{centering}
\caption{Meshed grid of four converter stations considered for the simulation
tests.\label{fig:test_grid_graph}}
\end{figure}

The considered test grid is a four station meshed grid, represented
in Figure~\ref{fig:test_grid_graph}, implemented in the EMT software.
Lines $e_{1,2}$ and $e_{1,3}$ are overhead lines. Lines $e_{1,4}$
and $e_{2,4}$ are hybrid lines comprising sections of underground
cables and overhead lines. Each transmission line is protected by
two relays located at its extremities. The EMT simulations are performed
at a sampling frequency $f_{\text{s}}=1\,\text{MHz}$ which also corresponds
to the frequency of the measurements. 

The MMC stations are simulated with parameters from Table~\ref{tab:MMC-parameters}
and the corresponding equivalent RLC approximation used in the parametric
model is given in Table~\ref{tab:MMC-parameters-eq}. The underground
cable and overhead line characteristics are displayed in Tables~\ref{tab:Underground-cable-characteristic}
and \ref{tab:OHL-charac} respectively. The corresponding distributed
parameters employed in the parametric model are in Table~\ref{tab:distributed_param}.

\begin{table}[h]
\caption{Characteristics of the MMC stations used for the EMT simulations \label{tab:MMC-parameters}}

\centering{}%
\begin{tabular}{lr}
\hline 
Rated power (MVA) & 1000\tabularnewline
DC rated voltage (kV) & 320\tabularnewline
Arm inductance (p.u.) & 0.15\tabularnewline
Capacitor energy in each submodule (kJ/MVA)  & 40\tabularnewline
Conduction losses of each IGBT/diode ($\Omega$) & 0.001\tabularnewline
Number of sub-modules per arm & 400\tabularnewline
\hline 
\end{tabular}
\end{table}

\begin{table}[tbh]
\centering{}\caption{Underground cable characteristics for the EMT simulations \label{tab:Underground-cable-characteristic}}
\begin{tabular}{ccc}
\hline 
 & Core & Screen\tabularnewline
\hline 
\hline 
Vertical distance (m) & \multicolumn{2}{c}{1.33}\tabularnewline
Outer radius (mm) & \multicolumn{2}{c}{63.9}\tabularnewline
Inside radius (mm) & 0 & 56.9\tabularnewline
Outside radius (mm) & 32 & 58.2\tabularnewline
Resistivity (n$\Omega$m) & 17.2 & 28.3\tabularnewline
\hline 
\end{tabular}
\end{table}

\begin{table}[h]
\caption{Overhead-line characteristics for the EMT simulations \label{tab:OHL-charac}}

\centering{}%
\begin{tabular}{lr||r||r}
\hline 
 & \multicolumn{3}{r}{}\tabularnewline
\hline 
\hline 
DC resistance (m$\Omega$/km) & \multicolumn{3}{c}{24}\tabularnewline
Outside diameter (cm) & \multicolumn{3}{c}{4.775}\tabularnewline
Horizontal distance (m) & \multicolumn{3}{c}{5}\tabularnewline
Vertical height at tower (m) & \multicolumn{3}{c}{30}\tabularnewline
Vertical height at mid-span (m) & \multicolumn{3}{c}{10}\tabularnewline
Soil resistivity ($\Omega$m) & \multicolumn{3}{c}{100}\tabularnewline
\hline 
\end{tabular}
\end{table}

\begin{table}[h]
\caption{Equivalent parameters of the MMC stations used in the parametric model\label{tab:MMC-parameters-eq}}

\centering{}%
\begin{tabular}{lr}
\hline 
Equivalent inductance (mH) & 8.1\tabularnewline
Equivalent resistance ($\Omega$) & 0.4\tabularnewline
Equivalent capacitance ($\mu$F) & 391\tabularnewline
\hline 
\end{tabular}
\end{table}

\begin{table}[h]
\caption{Transmission line distributed parameters at $1\,\text{kHz}$ used
in the parametric model \label{tab:distributed_param}}

\centering{}%
\begin{tabular}{llr||r||r||r||r||r||r||r||r}
\hline 
 & Underground cable & \multicolumn{9}{r}{Overhead lines}\tabularnewline
\hline 
\hline 
Series resistance $R$ (m$\Omega$/km) & 102 & \multicolumn{9}{c}{872}\tabularnewline
Series inductance $L$ (mH/km) & 0.123 & \multicolumn{9}{c}{1.84}\tabularnewline
Shunt capacitance $C$ (nF/km) & 241 & \multicolumn{9}{c}{7.68}\tabularnewline
Shunt conductance $G$ (nS/km) & -0.4 & \multicolumn{9}{c}{0.2}\tabularnewline
\hline 
\end{tabular}
\end{table}

\subsection{Modeling results\label{subsec:Modeling-results}}

This section presents the modeling results of the approach proposed
in Section~\ref{sec:Systematic-fault-modeling} and compares them
with EMT simulations. Faults in an underground cable section as well
as in an aerial part are both investigated. 

\subsubsection{Fault in an overhead line section}

Consider the fault $F_{1}$ in Figure~\ref{fig:test_grid_graph}
affecting the line between stations $q_{1}$ and $q_{4}$, on the
edge $e_{1,5}$ corresponding to an overhead line section. The fault
is located at a distance $d_{\text{f}}=100\,\text{km}$ from the station
$q_{1}$ and has an impedance $R_{\text{f}}=5\,\Omega$. The model
of the evolution of the voltage and current at the relay $R_{14}$
monitoring this line, located at station $q_{1}$ is compared with
the EMT data in Figure~\ref{fig:cable_fault_fit}. The proposed model
presents a very good accuracy compared to the EMT simulations for
both the current and voltage. The norm of the error in voltage and
current are less than 3~kV and 40~A respectively for the considered
observation window.

\begin{figure}[tbh]
\includegraphics[width=0.5\columnwidth]{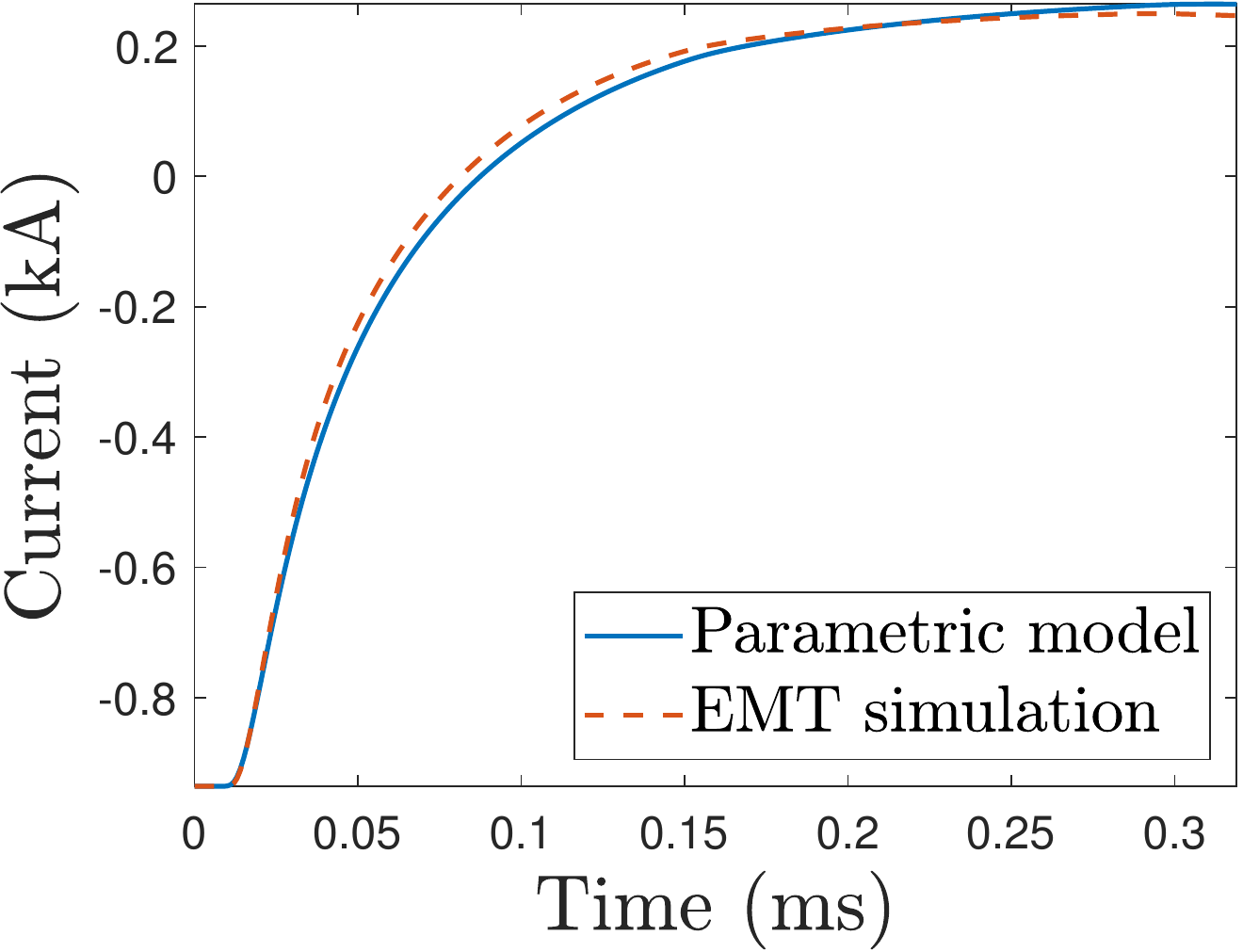}\includegraphics[width=0.5\columnwidth]{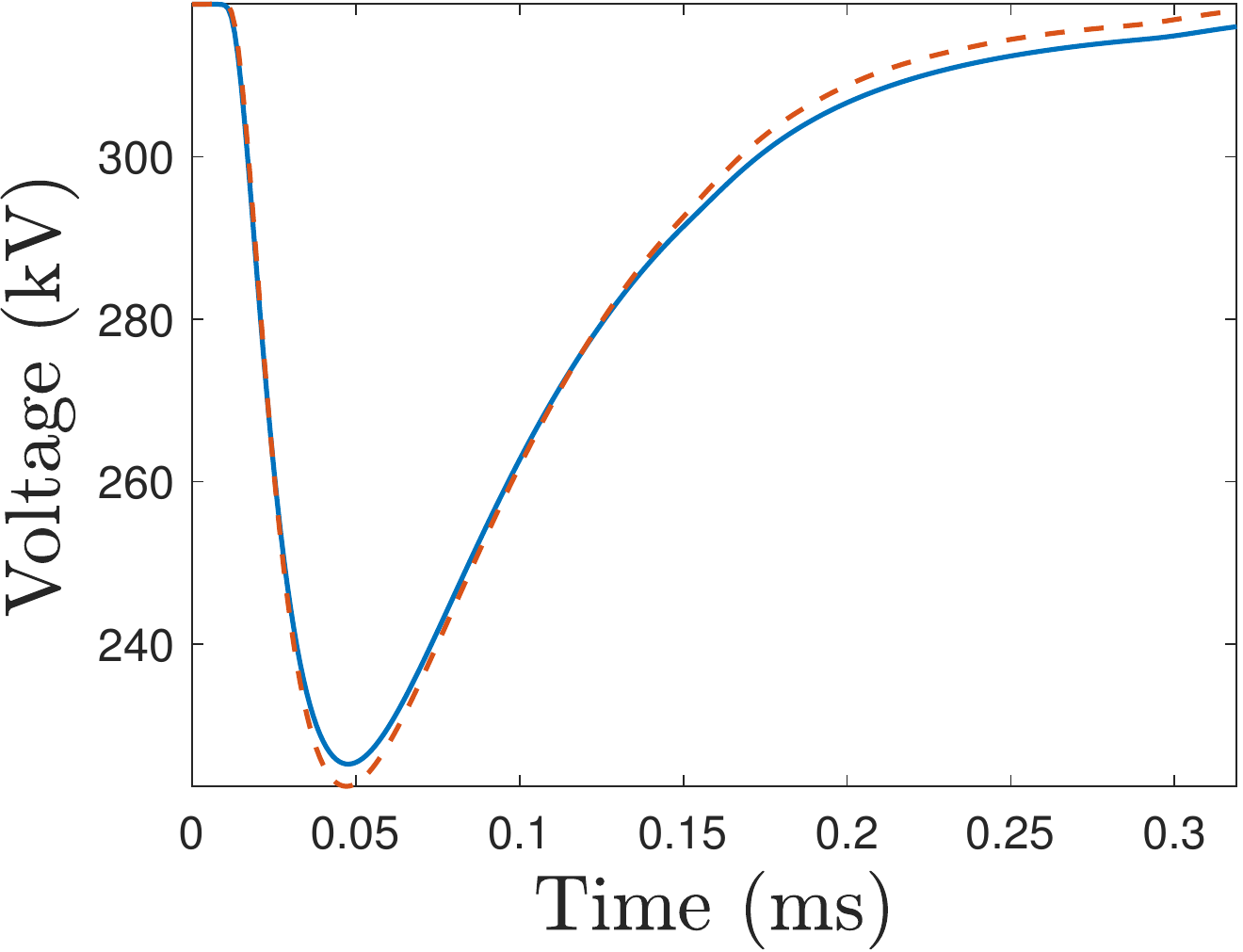}

\caption{Current and voltage simulation for overhead line fault occurring at
$d_{\text{f}}=100\,\text{km}$ from the station $q_{1}$, with a resistance
of $R_{\text{f}}=5\,\Omega$, as seen from relay $R_{14}$.\label{fig:line_fault_fit}}
\end{figure}

\subsubsection{Fault in underground cable section}

\begin{figure}[tbh]
\includegraphics[width=0.5\columnwidth]{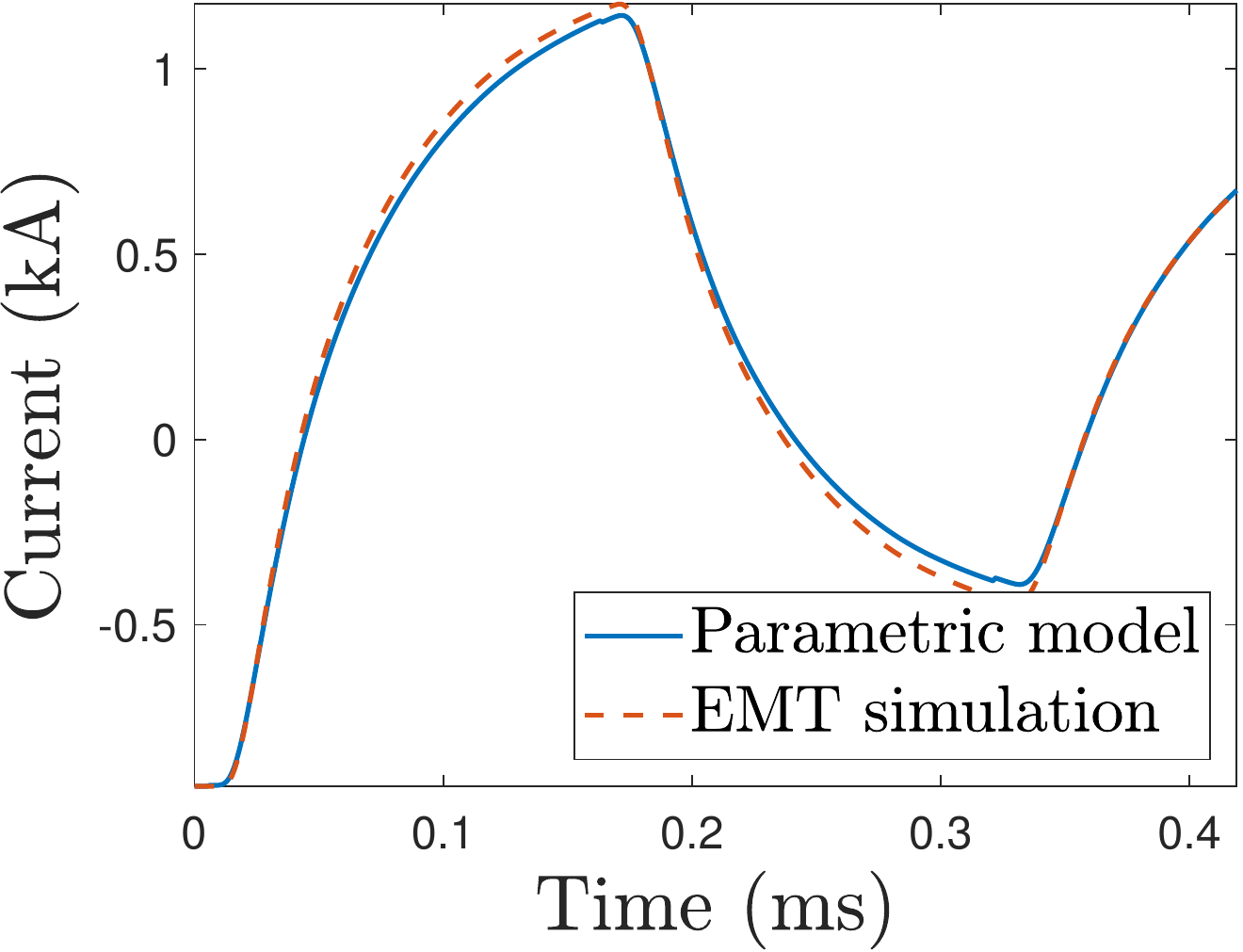}\includegraphics[width=0.5\columnwidth]{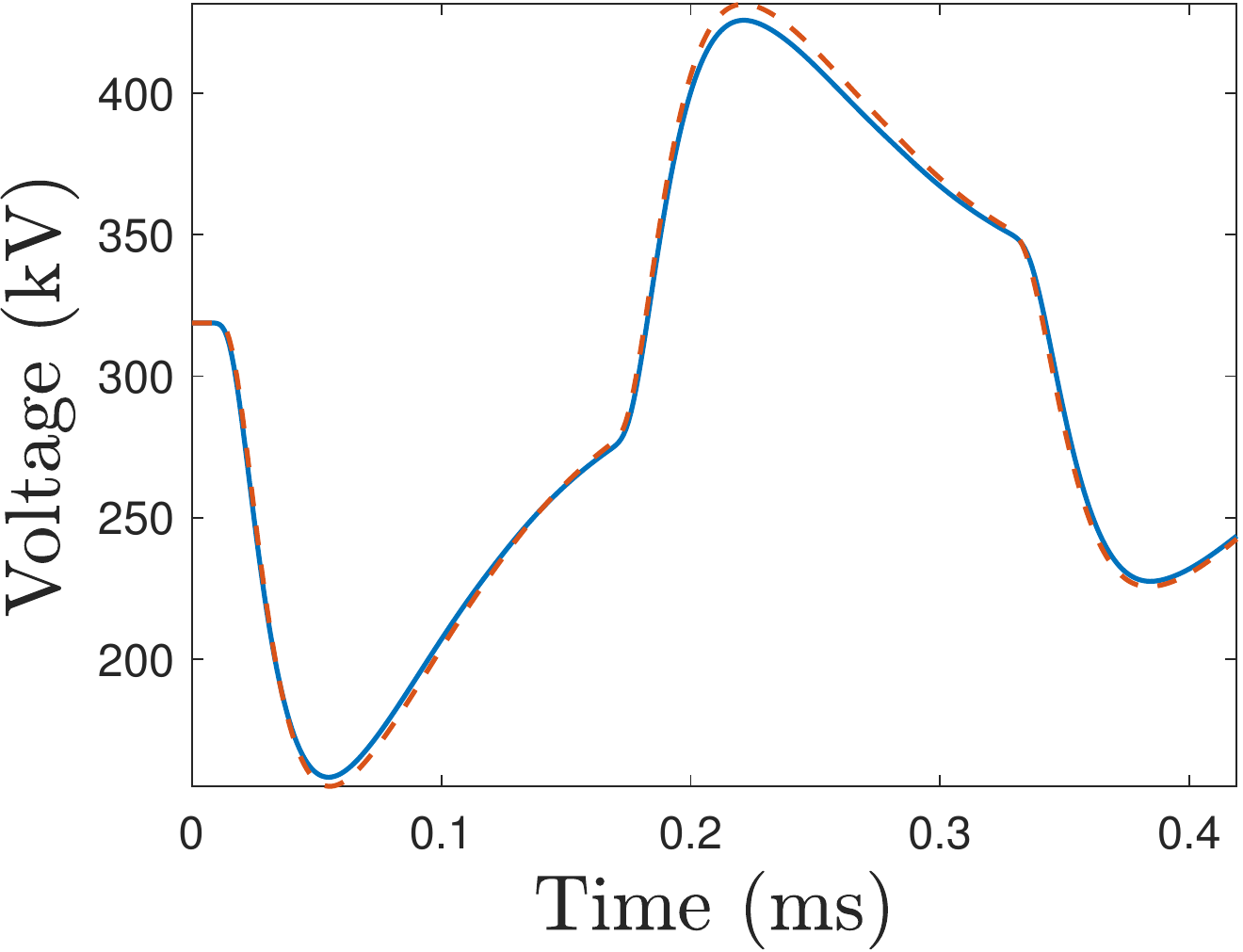}

\caption{Current (left) and voltage (right) simulation for cable fault occurring
at $d_{\text{f}}=135\,\text{km}$ from station 1, with resistance
of $R_{\text{f}}=0.1\,\Omega$, as seen from relay $R_{14}$.\label{fig:cable_fault_fit}}
\end{figure}

Consider the fault $F_{2}$ in Figure~\ref{fig:test_grid_graph}
affecting the line between stations $q_{1}$ and $q_{4}$, on the
edge $e_{5,6}$ corresponding to an underground section. The fault
is located at a distance $d_{\text{f}}=15\,\text{km}$ from the junction
$q_{5}$ and has an impedance of $R_{\text{f}}=0.1\,\Omega$. The
obtained model of the evolution of the voltage and current at the
relays $R_{14}$ and $R_{41}$ monitoring this line is compared with
the EMT simulation result in Figure~\ref{fig:cable_fault_fit}. The
norm of the error in voltage and current are less than 6~kV and 50~A
respectively for the considered observation window. The fault located
$15\,\text{km}$ away from $q_{5}$ and $5\,\text{km}$ from $q_{6}$
results in 3 significant TWs that must be taken into account in the
model (the small fault resistance makes the waves reflected at the
junction $q_{6}$ negligible). 

The model proposed in Section~\ref{sec:Systematic-fault-modeling}
is thus able to accurately represent the current and voltage TWs for
both underground and overhead line faults.

\subsection{Fault localization example \label{subsec:Illustrative-example}}

Consider the fault $F_{3}$ in Figure~\ref{fig:test_grid_graph}
affecting the line between stations $q_{1}$ and $q_{4}$, on the
edge $e_{6,7}$ corresponding to an overhead section. The fault is
located at $d_{\text{f}}=50\,\text{km}$ from the node $q_{6}$ and
has an impedance of $R_{\text{f}}=70\,\Omega$. The behavior of the
fault identification approach at the relay $R_{14}$ is analyzed.
In the least-squares criterion \eqref{eq:cost_func} the voltage and
current variances are set such that $\frac{\sigma_{i}^{2}}{\sigma_{v}^{2}}=4$.

Four algorithms are launched in parallel, each corresponding to a
different hypothesis relative to the faulty segment. Each algorithm
performs similarly: one iteration is performed in the minimization
of the cost function \eqref{eq:cost_func} every $\Delta n=10$ available
new measurements. 

\begin{figure}[tbh]
\includegraphics[width=0.5\columnwidth]{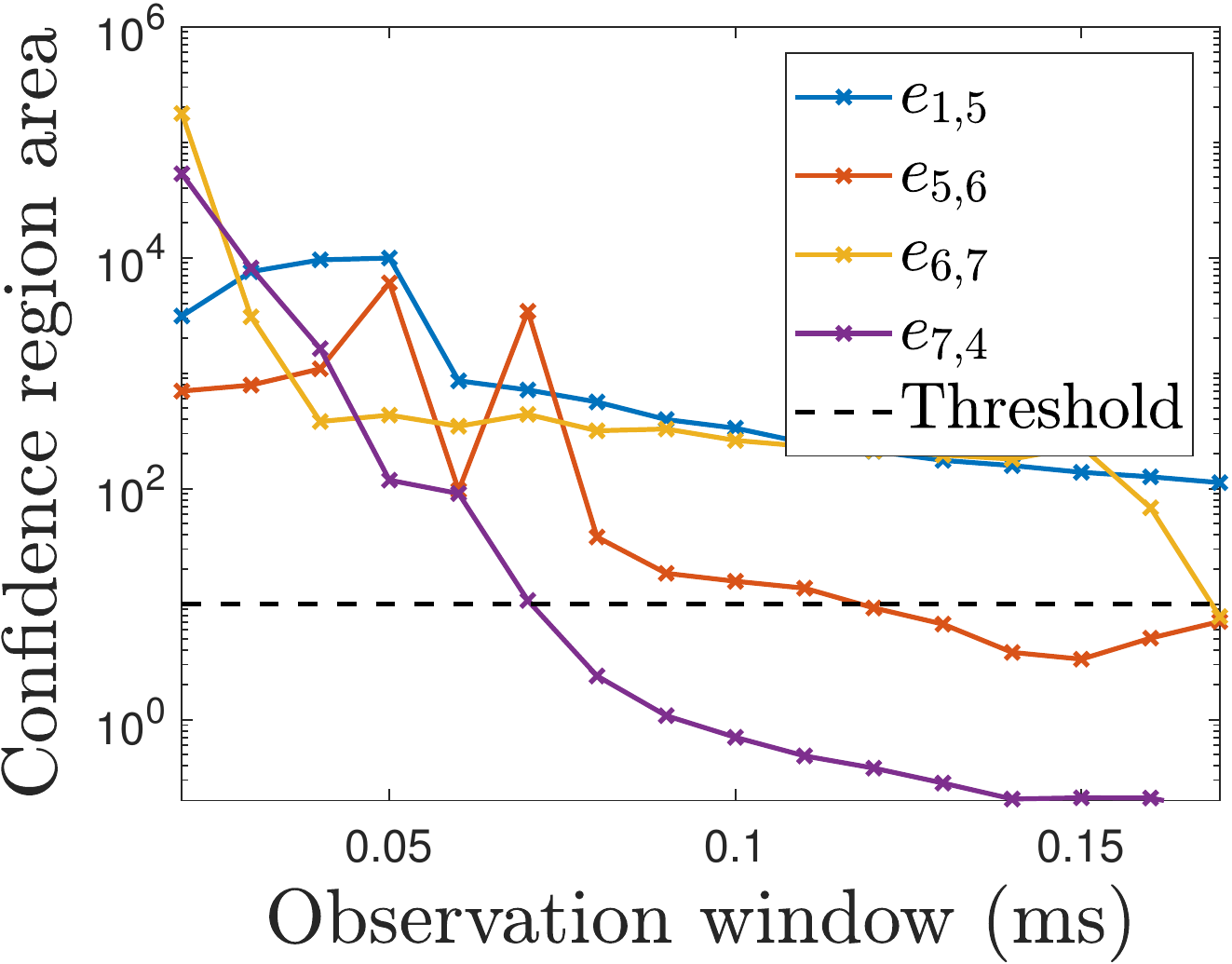}\includegraphics[width=0.5\columnwidth]{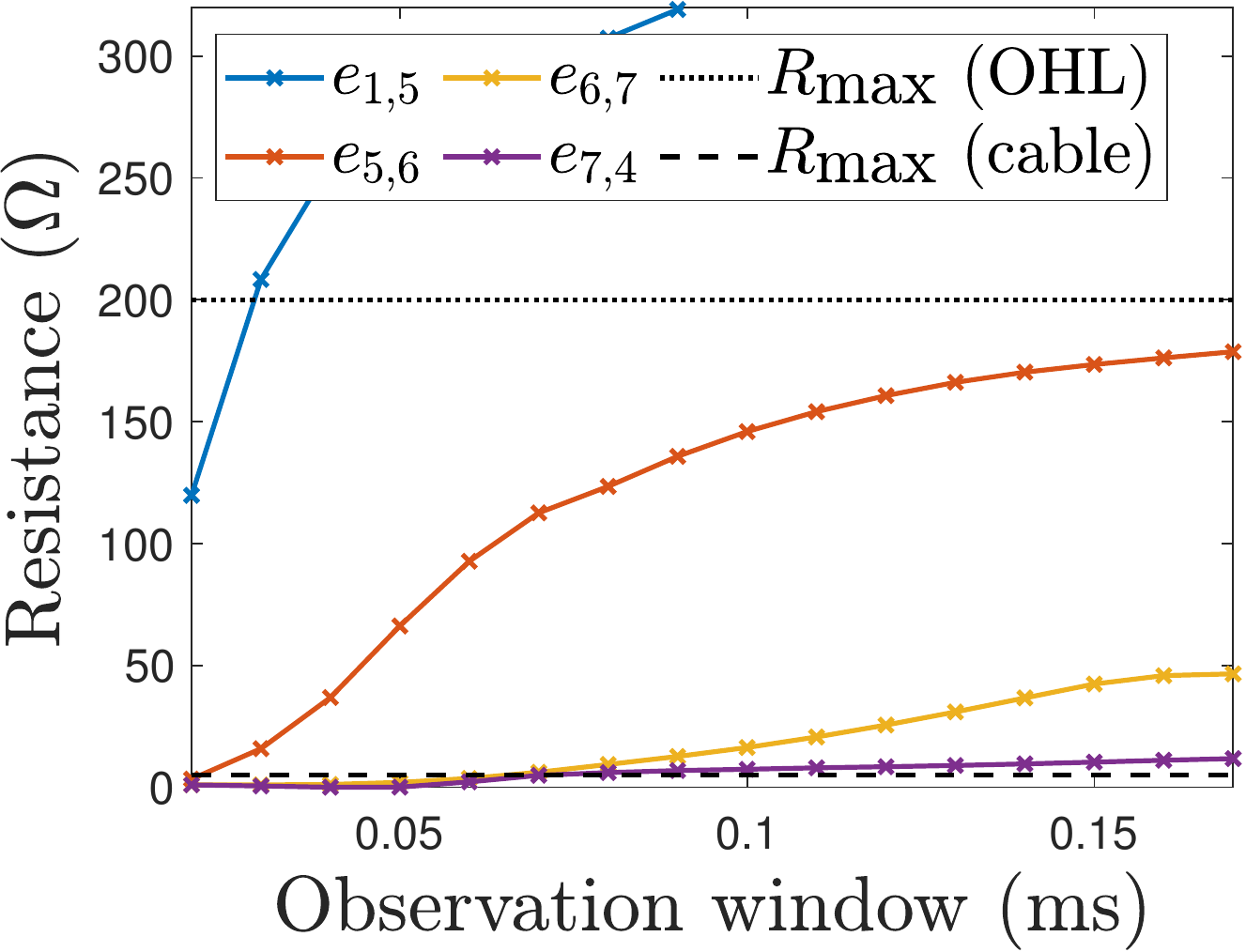}

\caption{Evolution of the area of the 95\% confidence ellipse (left) and value
of the estimated fault resistance (right) for each hypothesis.\label{fig:Evolution-Area_Cost}}
\end{figure}

The area of the 95\% confidence region for the estimated parameters
for the four different hypotheses are plotted in Figure~\ref{fig:Evolution-Area_Cost}
(left) and compared with the predetermined threshold $t_{95}=10$.
Three different hypotheses satisfy the accuracy test as the area of
their confidence region goes below the threshold. Nevertheless, the
validity test is not satisfied for the hypothesizes corresponding
to the faulty edges $e_{5,6}$ and $e_{7,8}$, as their estimated
fault resistances are above $R_{\text{max}}=5\,\Omega$ for the cables,
see Figure~\ref{fig:Evolution-Area_Cost} (right). Considering the
assumption that $e_{6,7}$ is the faulty edge, the estimated fault
parameters satisfy the validity test as the estimated resistance stays
below the maximum fault resistance $R_{\text{max}}=200\,\Omega$ for
an overhead section fault. 

Thus, the fault is correctly identified after 16 iterations on the
edge $e_{6,7}$ when the area of the confidence region for this hypothesis
goes below the threshold, see Section~\ref{subsec:Fault-identification}.
The estimated fault parameters after considering a measurement window
of $160\,\mu s$ are $\widehat{R}_{\text{f}}=47\,\Omega,\ \widehat{d}_{\text{f}}=54\,\text{km}$.

In this case, the localization of the fault in an overhead segment
indicates it is probably non-permanent and a reclosing of the line
may be attempted after some time.

\begin{figure}[tbh]
\includegraphics[width=0.5\columnwidth]{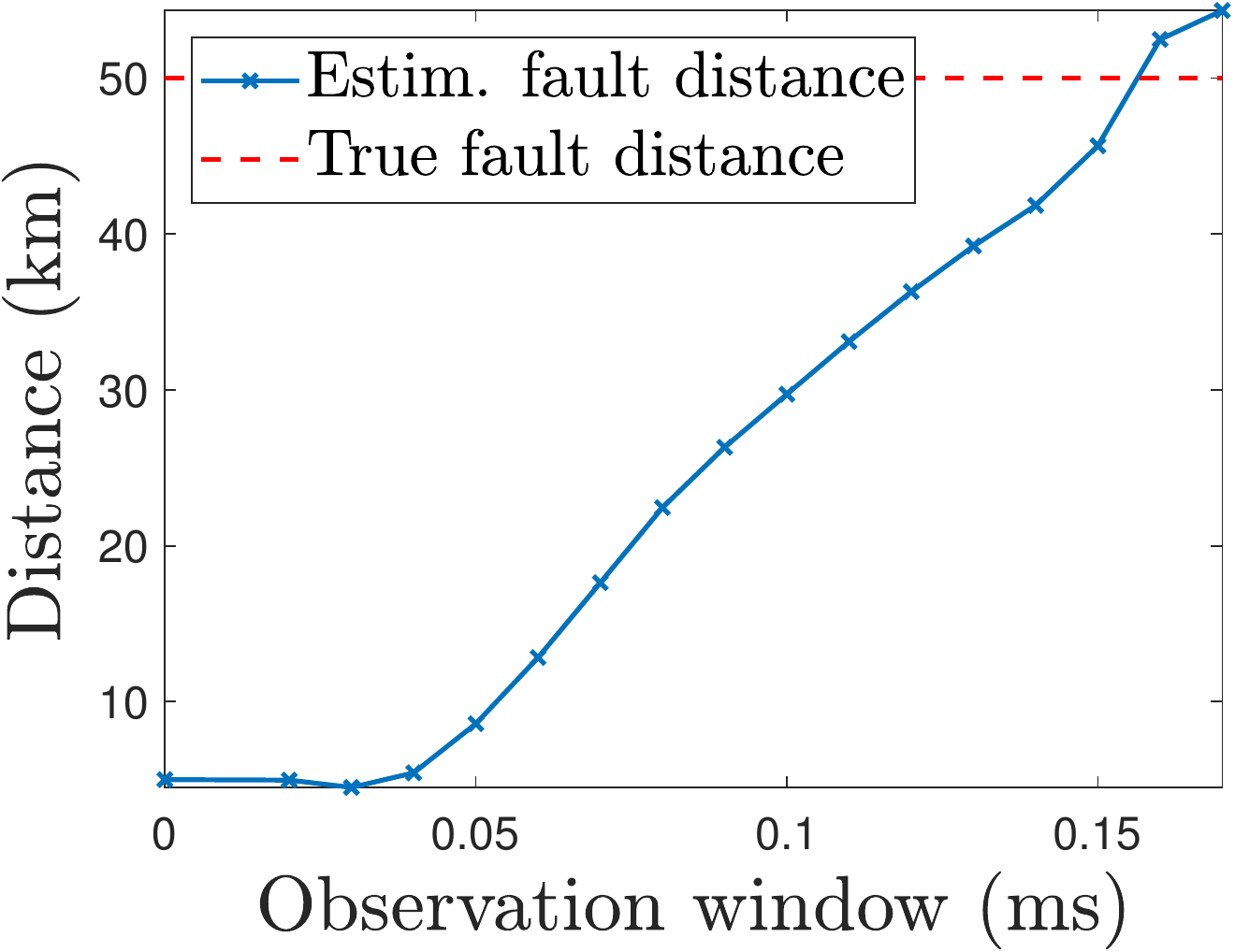}\includegraphics[width=0.5\columnwidth]{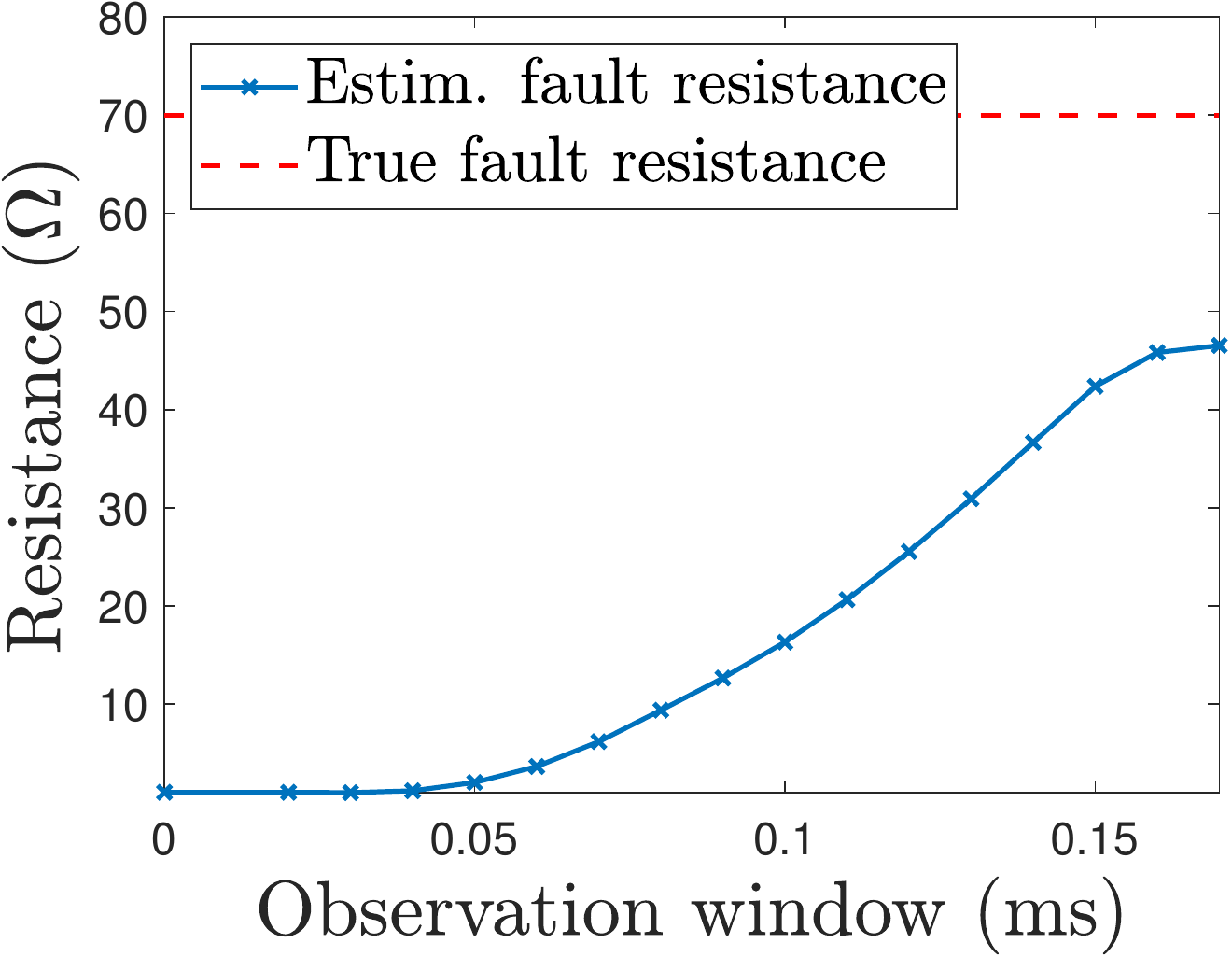}

\caption{Evolution with time of the estimated fault distance (left) and resistance
(right) considering the hypothesis of a fault in the edge $e_{6,7}$
(actual faulty edge). \label{fig:Estim_R_d}}
\end{figure}

Figure~\ref{fig:Estim_R_d} represents the evolution with the number
of iterations of the estimated fault distance and resistance considering
the hypothesis of a fault in the (actual faulty) edge $e_{6,7}$.

\begin{figure}[tbh]
\includegraphics[width=0.5\columnwidth]{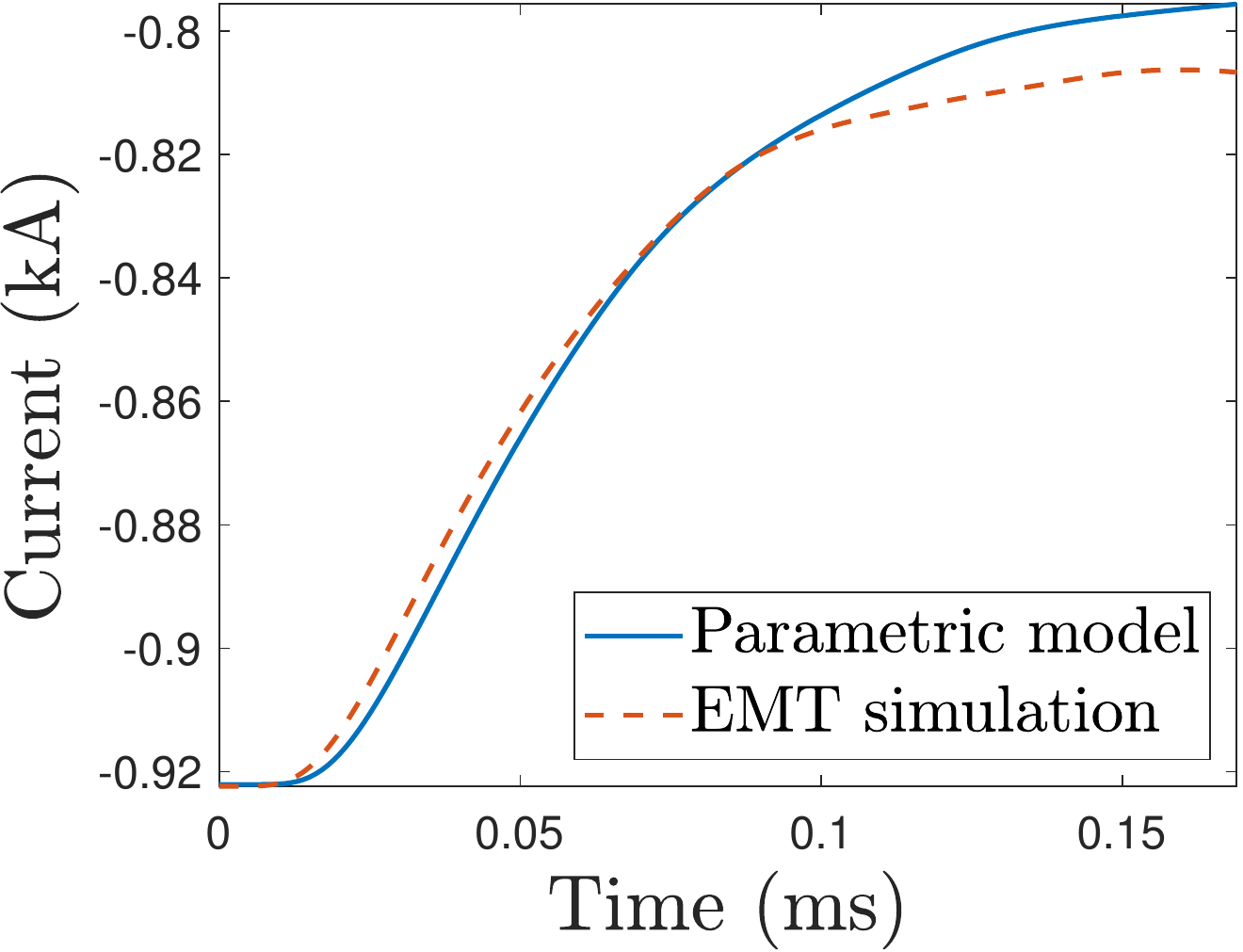}\includegraphics[width=0.5\columnwidth]{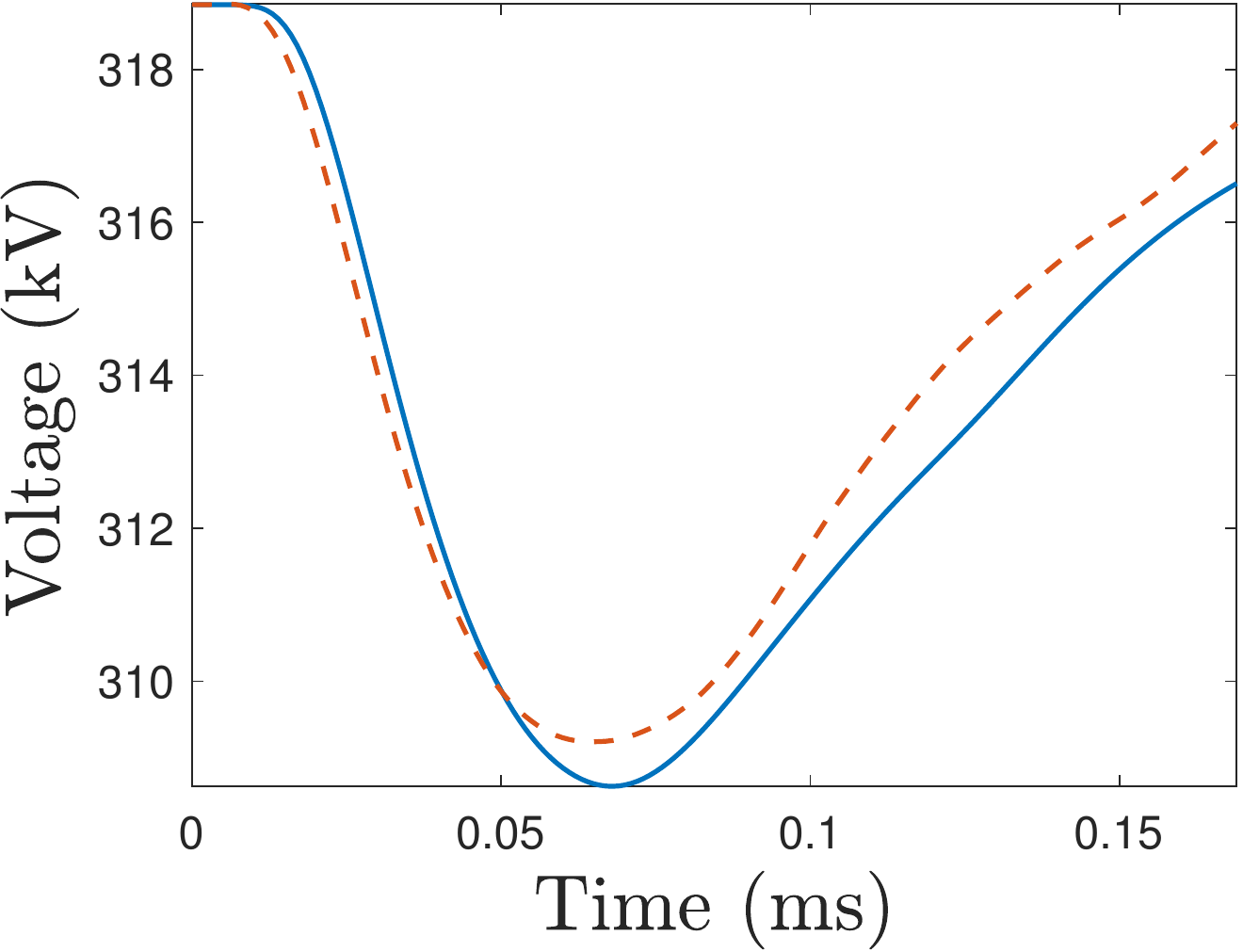}\caption{Comparison of the modeled and simulated voltage (right) and current
(left) at the relay $R_{14}$. The fault parameters used in the parametric
model are the ones obtained after 16 iterations: $\widehat{R}_{\text{f}}=47\,\Omega$
and $\widehat{d}_{\text{f}}=54\,\text{km}$.\label{fig:FIT_curr_volt}}
\end{figure}

The waveform of the voltage and current for the EMT simulation and
parametric model with the estimated fault parameters are compared
in Figure~\ref{fig:FIT_curr_volt}. The difference between the model
and the EMT data is always less than 10~A for the current and 1~kV
for the voltage and is mostly related to the difference between the
estimated and actual fault resistance.

\section{Conclusion}

This paper addresses the problem of traveling wave modeling in mixed
HVDC lines consisting of both overhead and underground parts. Though
EMT tools give an accurate representation of the transient phenomenon
occurring in case of a fault, they are ill-suited for protection applications
that require the fast evaluation of the transients. A model that
describes the transient behavior of the grid is proposed for single
conductor overhead lines and underground cables. A representation
of the grid and its components as a graph is considered. This allows
one to formally describe the multiple traveling waves generated after
the fault occurrence due to the reflections and transmissions occurring
at each junction within the grid. The obtained model depends explicitly
on the grid parameters as well as on the parameters of the fault such
as the fault distance and resistance.\\
When a fault is suspected, the model can be employed for the identification
and localization of the faulty segment of the line based on the estimation
of the fault parameters. The reclosing of the faulty line can then
be attempted if the line affects an overhead line where faults are
generally temporary. 


\onecolumn

\appendix{}

In this section the computation of the partial derivatives of the
voltage with respect to the fault distance and fault resistance are
established in a general case. The obtained expressions can be employed
in place of a finite difference approach for gradient evaluations
to reduce the computational burden of the parameter estimation algorithm
introduced in Section~\ref{subsec:Fault-parameter-estimation}. 

A fault is assumed to occur on an edge $e_{\text{f}}$ between nodes
$q_{k}$ and $q_{\ell}$ within a grid. The two edges connected to
the fault node $q_{\text{f}}$ are denoted as $e_{\text{f},k}$ and
$e_{\text{f},\ell}$ and are of lengths $d_{\text{f},k}$ and $d_{\text{f},\ell}$,
respectively. Assuming, without loss of generality, that $k<\ell$,
according to the fault distance convention \ref{eq:df_convention}:
$d_{\text{f}}=d_{\text{f},k}$. The computations are detailed for
a wave traveling though a path $\pi=\left(q_{n_{1}},\dots,q_{n_{m}}\right)$
where $q_{n_{i}}\in\mathcal{Q}_{\text{f}},\ i=1,\dots,m$ and $q_{n_{1}}=q_{\text{f}}$.

\subsection{Partial derivative with respect to the fault distance \label{subsec:Partial-derivative-dist}}

According to \eqref{eq:phy_model_temp} and \eqref{eq:model_v_pi_m},
the model of the voltage observed at node $q_{n_{m}}$ resulting from
a wave that traveled through the path $\pi$ can be expressed in the
times domain as
\[
v_{\pi}^{\text{m}}\left(\mathbf{p},k\right)=v_{\pi,\text{j}}\left(R_{\text{f}},\left(k-f_{\text{s}}\tau\left(d_{\text{f}}\right)\right)\right)\varotimes h_{\pi}\left(d_{\text{f}},k\right),
\]
where $\tau$ corresponds to the total propagation time through the
path $\pi$. The delay $\tau$ only depends on the fault distance
$d_{\text{f}}$.

The derivative with respect to the fault distance $d_{\text{f}}$
is then given by \eqref{eq:dv_dx_1}.
\begin{align}
\frac{\partial v_{\pi}^{\text{m}}\left(\mathbf{p},k\right)}{\partial d_{\text{f}}} & =v_{\pi,\text{j}}\left(R_{\text{f}},k-f_{\text{s}}\tau\left(d_{\text{f}}\right)\right)\varotimes\frac{\partial\left[h_{\pi}\left(d_{\text{f}},k\right)\right]}{\partial d_{\text{f}}}\label{eq:dv_dx_1}\\
 & +h_{\pi}\left(d_{\text{f}},k\right)\varotimes\frac{\partial\left[v_{\pi,\text{j}}^{\text{}}\left(R_{\text{f}},k-f_{\text{s}}\tau\left(d_{\text{f}}\right)\right)\right]}{\partial d_{\text{f}}}\nonumber \\
 & =v_{\pi,\text{j}}^{\text{}}\left(R_{\text{f}},k-f_{\text{s}}\tau\left(d_{\text{f}}\right)\right)\varotimes\frac{\partial\left[h_{\pi}\left(d_{\text{f}},k\right)\right]}{\partial d_{\text{f}}}\nonumber \\
 & -h_{\pi}\left(d_{\text{f}},k\right)\varotimes f_{\text{s}}\frac{\partial\tau}{\partial d_{\text{f}}}\frac{\partial\left[v_{\pi,\text{j}}^{\text{}}\left(R_{\text{f}},k-f_{\text{s}}\tau\left(d_{\text{f}}\right)\right)\right]}{\partial\left(k-f_{\text{s}}\tau\left(d_{\text{f}}\right)\right)}.\nonumber 
\end{align}
The delay $\tau$ due to the propagation along the path $\pi$ can
be expended as 

\begin{equation}
\tau\left(d_{\text{f}}\right)=\!\sum_{\substack{i\\
q_{n_{i}}\neq q_{\text{f}}\\
q_{n_{i+1}}\neq q_{\text{f}}
}
}\hspace{-0.2cm}\tau_{e_{n_{i},n_{i+1}}}+\hspace{-0.2cm}\sum_{\substack{\substack{i\\
e_{n_{i},n_{i+1}}=e_{\text{f},k}
}
}
}\hspace{-0.2cm}\tau_{e_{n_{i},n_{i+1}}}+\hspace{-0.2cm}\sum_{\substack{\substack{\substack{i\\
e_{n_{i},n_{i+1}}=e_{\text{f},\ell}
}
}
}
}\hspace{-0.2cm}\tau_{e_{n_{i},n_{i+1}}},\label{eq:tau_decomposition}
\end{equation}
where we have isolated the delays due to propagation along the edges
$e_{\text{f},k}$ and $e_{\text{f},\ell}$. Introducing $m_{\text{f},k}$
and $m_{\text{f},\ell}$ as the number of times the wave traveled
through the two edges connected to the fault $e_{\text{f},k}$ and
$e_{\text{f},\ell}$, one gets
\[
\tau\left(d_{\text{f}}\right)=\sum_{\substack{i\\
q_{n_{i}}\neq q_{\text{f}}\\
q_{n_{i+1}}\neq q_{\text{f}}
}
}\tau_{e_{n_{i},n_{i+1}}}+m_{\text{f},k}\frac{d_{\text{f}}}{c_{e_{\text{f}}}}+m_{\text{f},\ell}\frac{d_{e_{\text{f}}}-d_{\text{f}}}{c_{e_{\text{f}}}}
\]
The first sum correspond to propagation times along edges not connected
to the faulty node. Hence, assuming the propagation speed does not
depend on the fault distance

\begin{equation}
\frac{\partial\tau}{\partial d_{\text{f}}}=\frac{m_{\text{f},k}-m_{\text{f},\ell}}{c_{e_{\text{f}}}}.\label{eq:tau_derived}
\end{equation}

Consider now the finite impulse response filter $h_{\pi}$ that represents
the total distortion along the considered path $\pi$. The filter
is expressed in the frequency domain in \eqref{eq:H_pi}, where the
same decomposition as in \eqref{eq:tau_decomposition} has been performed,
\begin{align}
H_{\pi}\left(d_{\text{f}},\omega\right)= & \hspace{-0.2cm}\prod_{\substack{i=1\\
q_{n_{i}}\neq q_{\text{f}}\\
q_{n_{i+1}}\neq q_{\text{f}}
}
}\hspace{-0.2cm}H_{(q_{n_{i}},q_{n_{i+1}})}\left(\omega\right)\times\hspace{-0.2cm}\prod_{\substack{i=1\\
(q_{n_{i}},q_{n_{i+1}})=e_{\text{f,}k}
}
}\hspace{-0.2cm}H_{(q_{n_{i}},q_{n_{i+1}})}\left(d_{\text{f}},\omega\right)\times\hspace{-0.2cm}\prod_{\substack{i=1\\
(q_{n_{i}},q_{n_{i+1}})=e_{\text{f, \ensuremath{\ell}}}
}
}\hspace{-0.2cm}H_{(q_{n_{i}},q_{n_{i+1}})}\left(d_{\text{f}},\omega\right)\label{eq:H_pi}\\
= & \prod_{\substack{i=1\\
q_{n_{i}}\neq q_{\text{f}}\\
q_{n_{i+1}},\neq q_{\text{f}}
}
}H_{(q_{n_{i}},q_{n_{i+1}})}\left(\omega\right)\times H_{e_{\text{f,}k}}^{m_{\text{f},k}}\left(d_{\text{f}},\omega\right)\times H_{e_{\text{f,}\ell}}^{m_{\text{f},\ell}}\left(d_{\text{f}},\omega\right).\nonumber 
\end{align}
Taking the derivative with respect to the fault distance, and omitting
the dependency in $\omega,$ one gets \eqref{eq:d_Hpi_dx} 
\begin{align}
\frac{\partial H_{\pi}\left(d_{\text{f}},\omega\right)}{\partial d_{\text{f}}} & =\prod_{\substack{i=1\\
q_{n_{i}}\neq q_{\text{f}}\\
q_{n_{i+1}}\neq q_{\text{f}}
}
}H_{(q_{n_{i}},q_{n_{i+1}})}\left[m_{\text{f},k}H_{e_{\text{f,}\ell}}^{m_{\text{f},\ell}}H_{e_{\text{f,}k}}^{m_{\text{f},k}-1}\frac{\partial H_{e_{\text{f,}k}}}{\partial d_{\text{f}}}+m_{\text{f},\ell}H_{e_{\text{f,}\ell}}^{m_{\text{f},\ell}-1}H_{e_{\text{f,}k}}^{m_{\text{f},k}}\frac{\partial H_{e_{\text{f,}\ell}}}{\partial d_{\text{f}}}\right]\label{eq:d_Hpi_dx}\\
 & =\underbrace{\prod_{\substack{i=1\\
q_{n_{i}}\neq q_{\text{f}}\\
q_{n_{i+1}}\neq q_{\text{f}}
}
}H_{(q_{n_{i}},q_{n_{i+1}})}H_{e_{\text{f,}k}}^{m_{\text{f},k}}H_{e_{\text{f,}k}}^{m_{\text{f},\ell}}}_{H_{\pi}}\left[m_{e_{\text{f},k}}H_{e_{\text{f,}k}}^{-1}\frac{\partial H_{e_{\text{f,}k}}}{\partial d_{\text{f}}}+m_{e_{\text{f},\ell}}H_{e_{\text{f,}\ell}}^{-1}\frac{\partial H_{e_{\text{f,}\ell}}}{\partial d_{\text{f}}}\right]\nonumber 
\end{align}
Moreover, we assume that a wave that travels successively through
$e_{\text{f},k}$ and $e_{\text{f},\ell}$ is prone to the same distortion
as a wave that travels through $e_{\text{f}}$, \textit{i.e.}, 

\[
H_{e_{\text{f,}k}}\left(d_{\text{f}},\omega\right)H_{e_{\text{f,}\ell}}\left(d_{\text{f}},\omega\right)=H_{e_{\text{f}}}\left(\omega\right)
\]
Where $H_{e_{\text{f}}}$ does not depend on the fault distance. Hence, 

\[
\frac{\partial H_{e_{\text{f,}\ell}}\left(d_{\text{f}},\omega\right)}{\partial d_{\text{f}}}=-\frac{\partial H_{e_{\text{f,}k}}\left(d_{\text{f}},\omega\right)}{\partial d_{\text{f}}}\frac{H_{e_{\text{f}}}\left(\omega\right)}{H_{e_{\text{f,}k}}\left(d_{\text{f}},\omega\right)^{2}}
\]
The expression of $\frac{\partial H_{\pi}\left(d_{\text{f}},\omega\right)}{\partial d_{\text{f}}}$
can thus be further simplified

\begin{align*}
\frac{\partial H_{\pi}(d_{\text{f}},\omega)}{\partial d_{\text{f}}} & =H_{\pi}\left[m_{e_{\text{f},k}}H_{e_{\text{f,}k}}^{-1}\frac{\partial H_{e_{\text{f,}k}}}{\partial d_{\text{f}}}+m_{e_{\text{f},\ell}}H_{e_{\text{f,}\ell}}^{-1}\frac{\partial H_{e_{\text{f,}\ell}}}{\partial d_{\text{f}}}\right]\\
 & =H_{\pi}\left[m_{e_{\text{f},k}}H_{e_{\text{f,}k}}^{-1}\frac{\partial H_{e_{\text{f,}k}}}{\partial d_{\text{f}}}-\frac{m_{e_{\text{f},\ell}}H_{e_{\text{f}}}}{H_{e_{\text{f,}\ell}}H_{e_{\text{f,}k}}^{2}}\frac{\partial H_{e_{\text{f,}k}}}{\partial d_{\text{f}}}\right]\\
 & =H_{\pi}\left[m_{e_{\text{f},k}}H_{e_{\text{f,}k}}^{-1}\frac{\partial H_{e_{\text{f,}k}}}{\partial d_{\text{f}}}-m_{e_{\text{f},\ell}}H_{e_{\text{f,}k}}^{-1}\frac{\partial H_{e_{\text{f,}k}}}{\partial d_{\text{f}}}\right]\\
 & =H_{\pi}H_{e_{\text{f,}k}}^{-1}\frac{\partial H_{e_{\text{f,}k}}}{\partial d_{\text{f}}}\left[m_{e_{\text{f},k}}-m_{e_{\text{f},\ell}}\right].
\end{align*}
Taking the inverse Fourier transform
\begin{equation}
\frac{\partial h_{\pi}\left(d_{\text{f}},k\right)}{\partial d_{\text{f}}}=\left[m_{e_{\text{f},k}}-m_{e_{\text{f},\ell}}\right]\mathcal{F}^{-1}\left\{ \frac{H_{\pi}\left(d_{\text{f}},\omega\right)}{H_{e_{\text{f,}k}}}\frac{\partial H_{e_{\text{f,}k}}}{\partial d_{\text{f}}}\right\} \label{eq:impulse_responsed_derived}
\end{equation}
The derivative of the impulse response $H_{e_{\text{f,}k}}=\mathcal{F}\left(h_{e_{\text{f,}k}}\right)$
can be obtained from \eqref{eq:impulse_he}
\[
h_{e}(k)=f_{\text{s}}\cdot\left(u_{d,\rho}(d_{\text{f}},k+1)-u_{d,\rho}(d_{\text{f}},k)\right)
\]
and the linear interpolation \eqref{eq:step_d_interp}, 

\[
u_{d,\rho}(d_{\text{f}},k)=\frac{u_{d_{2},\rho}(k)-u_{d_{1},\rho}(k)}{(d_{2}-d_{1})}(d_{\text{f}}-d_{1})+u_{d_{1},\rho}(k)
\]
where $d_{1}<d_{\text{f}}<d_{2}$. Hence, 

\[
\frac{\partial u_{d,\rho}(k)}{\partial d_{\text{f}}}=\frac{u_{d_{2},\rho}(k)-u_{d_{1},\rho}(k)}{(d_{2}-d_{1})}.
\]

The last derivative to compute in \eqref{eq:dv_dx_1} is 

\[
\frac{\partial v_{\pi,\text{j}}\left(R_{\text{f}},k-f_{\text{s}}\tau\left(d_{\text{f}}\right)\right)}{\partial\left(k-f_{\text{s}}\tau\left(d_{\text{f}}\right)\right)}=\left.\frac{\partial v_{\pi,\text{j}}\left(R_{\text{f}},k'\right)}{\partial k'}\right|_{k'=k-f_{\text{s}}\tau},
\]
which may be approximated by the finite difference 

\begin{equation}
\frac{\partial v_{\pi,\text{j}}\left(R_{\text{f}},k'\right)}{\partial k'}\simeq\left(v_{\pi,\text{j}}\left(R_{\text{f}},k'+1\right)-v_{\pi,\text{j}}\left(R_{\text{f}},k'\right)\right).\label{eq:dv_pi_j_dx}
\end{equation}
The final voltage derivative \pageref{eq:d_v_dx_tot} is obtained
combining \eqref{eq:tau_derived}, \eqref{eq:impulse_responsed_derived}
and \eqref{eq:dv_pi_j_dx}, 
\begin{equation}
\frac{\partial v_{\pi}^{\text{m}}\left(\mathbf{p},k\right)}{\partial d_{\text{f}}}=\left[v_{\pi,\text{j}}^{\text{}}\left(R_{\text{f}},k-f_{\text{s}}\tau\left(d_{\text{f}}\right)\right)\varotimes\frac{\partial\left[h_{\pi}\left(d_{\text{f}},k\right)\right]}{\partial d_{\text{f}}}-f_{\text{s}}\frac{\partial\tau}{\partial d_{\text{f}}}h_{\pi}\left(d_{\text{f}},k\right)\varotimes\frac{\partial\left[v_{\pi,\text{j}}^{\text{}}\left(R_{\text{f}},k-f_{\text{s}}\tau\left(d_{\text{f}}\right)\right)\right]}{\partial\left(k-f_{\text{s}}\tau\left(d_{\text{f}}\right)\right)}\right].\label{eq:d_v_dx_tot}
\end{equation}
 The evaluation of \eqref{eq:tau_derived} is simple as it only requires
counting how many times the wave traveled through the edges connected
to the fault. Similarly, \eqref{eq:dv_pi_j_dx} involves a discrete
differentiation. The computation of \eqref{eq:impulse_responsed_derived}
is more demanding but still relatively efficient as $h_{\pi}$ is
already available from the computations of the voltage. This approach
thus leads to a direct evaluation of the derivative with respect to
the fault distance more effective than a finite difference approach.

\subsection{Partial derivative with respect to the fault resistance\label{subsec:Partial-derivative-resist}}

The parametric model \eqref{eq:phy_model_temp}, \eqref{eq:model_v_pi_m}
depends on the fault resistance $R_{\text{f}}$ only through the interactions
at the fault location, appearing in the reflection and transmission
coefficients \eqref{eq:refl_coef} and \eqref{eq:trans_coef}. Furthermore,
in the loss\textendash less transmission line model, the surge impedance
is a real number. Since the fault impedance is considered as purely
resistive, the reflection and transmission coefficients at the fault
location are thus also real numbers. 

The part of the model that computes the reflection and transmission
at the different interfaces $V_{\pi,\text{j}}^{\text{}}\left(R_{\text{f}},s\right)$
is
\[
V_{\pi,\text{j}}^{\text{}}(R_{\text{f}},s)=\prod_{i=1}^{n}J_{e_{n_{i-1},n_{i}}\shortrightarrow q_{n_{i}}}(s,R_{\text{f}})\frac{\exp(-t_{\text{f}}s)}{s}V_{\text{bf}}
\]
where the first term accounts for the initial surge at the fault location
$J_{e_{n_{0},n_{1}}\shortrightarrow q_{n_{1}}}=K_{q_{_{n_{1}},},q_{_{n_{2}},}\hookleftarrow q_{\text{f}}}$.
One can separate the reflections and transmissions at the fault location,
which involves the fault resistance, and the interactions at the other
junctions
\begin{align*}
 & V_{\pi,\text{j}}^{\text{}}(R_{\text{f}},s)=e^{-st_{\text{f}}}V_{\text{bf}}K_{q_{_{n_{1}},},q_{_{n_{2}},}\hookleftarrow q_{\text{f}}}\left(R_{\text{f}}\right)\prod_{\substack{i\\
q^{i}\neq q_{\text{f}}
}
}J_{e_{n_{i-1},n_{i}}\shortrightarrow q_{n_{i}}}\\
 & \times\prod_{\substack{i=1\\
q_{n_{i}}=q_{\text{f}}\\
q_{n_{i+1}}=q_{n_{i-1}}
}
}\underbrace{J_{e_{n_{i-1},n_{i}}\shortrightarrow q_{n_{i}}}}_{=K_{q_{_{n_{1}},},q_{_{n_{2}},}\hookleftarrow q_{\text{f}}}}\times\prod_{\substack{i=1\\
q_{n_{i}}=q_{\text{f}}\\
q_{n_{i+1}}\neq q_{n_{i-1}}
}
}\underbrace{J_{e_{n_{i-1},n_{i}}\shortrightarrow q_{n_{i}}}}_{=T_{q_{_{n_{1}},},q_{_{n_{2}},}\rightarrow q_{\text{f}}}}.
\end{align*}
The reflection and transmission coefficients at the fault location
are noted $K_{\text{f}}$ and $T_{\text{f}}$ to lighten notations.
Consider the numbers $m_{\text{K}_{\text{f}}}$ and $m_{T_{\text{T}_{\text{f}}}}$
of the reflections and transmissions at the fault location. Hence, 

\[
V_{\pi,\text{j}}^{\text{}}(R_{\text{f}},s)=e^{-st_{\text{f}}}V_{\text{bf}}K_{\text{f}}^{1+m_{K_{\text{f}}}}\left(R_{\text{f}}\right)T_{\text{f}}^{m_{T_{\text{f}}}}\left(R_{\text{f}}\right)\prod_{\substack{i\\
q_{n_{i}}\neq q_{\text{f}}
}
}J_{e_{n_{i-1},n_{i}}\shortrightarrow q_{n_{i}}}.
\]
The derivative with respect to the fault resistance is detailed in
\eqref{eq:dv_pi_dR}
\begin{align}
\frac{\partial V_{\pi,\text{j}}^{\text{}}(R_{\text{f}},s)}{\partial R_{\text{f}}} & =\exp(-st_{\text{f}})V_{\text{bf}}\prod_{\substack{i\\
q^{i}\neq q_{\text{f}}
}
}J_{e_{n_{i-1},n_{i}}\shortrightarrow q_{n_{i}}}\label{eq:dv_pi_dR}\\
\times & \left[\left(1+m_{K_{\text{f}}}\right)K_{\text{f}}^{m_{K_{\text{f}}}}\frac{\partial K_{e_{\text{f}\shortrightarrow f}}}{\partial R_{\text{f}}}\times T_{\text{f}}^{m_{T_{\text{f}}}}+m_{T_{\text{f}}}T_{\text{f}}^{m_{K_{\text{f}}}-1}\frac{\partial T_{e_{\text{f}\shortrightarrow f}}}{\partial R_{\text{f}}}\times K_{\text{f}}^{1+m_{T_{\text{f}}}}\right]\nonumber \\
 & =\underbrace{\exp(-st_{\text{f}})V_{\text{bf}}\ast\prod_{\substack{i\\
q^{i}\neq q_{\text{f}}
}
}J_{e_{n_{i-1},n_{i}}\shortrightarrow q_{n_{i}}}\times K_{\text{f}}^{1+m_{K_{\text{f}}}}\times T_{\text{f}}^{m_{T_{\text{f}}}}}_{=V_{\pi,\text{j}}^{\text{}}(R_{\text{f}},s)}\nonumber \\
\times & \left[\left(1+m_{K_{\text{f}}}\right)K_{\text{f}}^{-1}\frac{\partial K_{\text{f}}}{\partial R_{\text{f}}}+m_{T_{\text{f}}}T_{\text{f}}^{-1}\frac{\partial T_{\text{f}}}{\partial R_{\text{f}}}\right].\nonumber 
\end{align}
Hence, one gets

\[
\frac{\partial V_{\pi,\text{j}}^{\text{}}(R_{\text{f}},s)}{\partial R_{\text{f}}}=V_{\pi,\text{j}}^{\text{}}(R_{\text{f}},s)\left[\left(1+m_{K_{\text{f}}}\right)K_{\text{f}}^{-1}\frac{\partial K_{\text{f}}}{\partial R_{\text{f}}}+m_{T_{\text{f}}}T_{\text{f }}^{-1}\frac{\partial T_{\text{f}}}{\partial R_{\text{f}}}\right]
\]
The reflection and transmission at the fault location are 
\begin{align*}
K_{\text{f}}= & \frac{R_{\text{f}}}{0.5Z_{\text{s},e_{\text{f}}}+R_{\text{f}}}\\
T_{\text{f}}= & \frac{-Z_{\text{s},e_{\text{f}}}}{Z_{\text{s},e_{\text{f}}}+2R_{\text{f}}}
\end{align*}
whose derivatives are 
\begin{align*}
\frac{\partial T_{\text{f}}}{\partial R_{\text{f}}}= & \frac{0.5Z_{e_{\text{f}}}}{\left(0.5Z_{e_{\text{f}}}+R_{\text{f}}\right)^{2}}=\frac{K_{\text{f}}T_{\text{f}}}{R_{\text{f}}}
\end{align*}

\begin{align*}
\frac{\partial K_{\text{f}}}{\partial R_{\text{f}}} & =\frac{\partial\left(T_{\text{f}}-1\right)}{\partial R_{\text{f}}}=\frac{\partial T_{\text{f}}}{\partial R_{\text{f}}}
\end{align*}
Hence the previous expressions can be even further simplified
\[
\frac{\partial V_{\pi,\text{j}}^{\text{}}(R_{\text{f}},s)}{\partial R_{\text{f}}}=V_{\pi,\text{j}}^{\text{}}(R_{\text{f}},s)\frac{\left[(1+m_{K_{\text{f}}})T_{\text{f}}+m_{T_{\text{f}}}K_{\text{f}}\right]}{R_{_{\text{f}}}}.
\]
The voltage derivative with respect to the fault resistance hence
amounts to a multiplication by a real coefficient of the voltage expression.
Thus the same coefficient can be applied in time domain, \textit{i.e.}, 

\begin{equation}
\frac{\partial v_{\pi}^{\text{m}}\left(\mathbf{p},t_{k}\right)}{\partial R_{\text{f}}}=v_{\pi}^{\text{m}}\left(\mathbf{p},t_{k}\right)\frac{\left[(1+m_{K_{\text{f}}})T_{\text{f}}+m_{T_{\text{f}}}K_{\text{f}}\right]}{R_{_{\text{f}}}}.
\end{equation}

\end{document}